\documentclass[journal]{IEEEtran}
\usepackage{mathrsfs}
\usepackage{amssymb,amsmath}
\usepackage{amsfonts}
\usepackage{graphicx}
\usepackage{multirow}
\usepackage[all]{xy}

\newtheorem{theorem}{Theorem}
\newtheorem{definition}{Definition}
\newtheorem{lemma}{Lemma}
\newtheorem{corollary}{Corollary}

\newtheorem{myclaim}{Claim}

\newcommand{\from}{\leftarrow}
\newcommand{\frametext}[2]
{
\indent
\\
\refstepcounter{figure}
\fbox{
\parbox[t]{8.3cm}
{#1
\begin{center}
{\small \textbf{Fig. \arabic{figure}.} #2}
\end{center}
}
}
\vspace{0.2cm}
}

\begin{document}

\title{On Optimal Secure Message Transmission by Public Discussion}
\author{Hongsong Shi,
Shaoquan Jiang,
Reihaneh Safavi-Naini,
\and Mohammed Ashraful Tuhin
\thanks{Manuscript received September 30, 2009.}
\thanks{H. Shi is with the School of Computer Science and
Technology, University of Electronic Science and Technology of
China, ChengDu, China 610054 \& with the Department of Computer
Science University of Calgary, Calgary,  Canada T2N 1N4 }
\thanks{S. Jiang is with the School of Computer Science and
Technology, University of Electronic Science and Technology of
China, ChengDu, China 610054}
\thanks{R. Safavi-Naini is with the Department of Computer Science,
University of Calgary, Calgary,  Canada T2N 1N4 }
\thanks{M. A. Tuhin is with the Department of Computer Science University of
Calgary, Calgary,  Canada T2N 1N4 }

}

\maketitle

\begin{abstract}
In a secure message transmission (SMT) scenario a sender wants to
send a message in a private and reliable way to a receiver. Sender
and receiver are connected by $n$ vertex disjoint paths, referred to
as wires, $t$ of which can be controlled by an adaptive adversary
with unlimited computational resources.  In Eurocrypt 2008, Garay
and Ostrovsky considered an SMT scenario where sender and receiver
have access to a public discussion channel and showed that secure
and reliable communication is possible when $n \geq t+1$. In this
paper we will show that a secure protocol requires at least 3 rounds
of communication and 2 rounds invocation of the public channel and
hence give a complete answer to the open question raised by Garay
and Ostrovsky.  We also describe a round optimal protocol that has
\emph{constant} transmission rate over the public channel.
\end{abstract}


\section{Introduction}
\IEEEPARstart{D}olev, Dwork, Waarts and Yung \cite{DDWY93}
introduced \emph{Secure Message Transmission} (SMT) systems to
address the problem of delivering a message from sender $\cal{S}$ to
receiver $\cal R$ in a network guaranteeing {\em reliability} and
{\em privacy}. $\cal{S}$ is connected to $\cal R$ by $n$ node
disjoint paths, referred to as {\em wires}, $t$ controlled by the
adversary with unlimited computational power.

A {\em perfectly} secure message transmission \emph{or} {PSMT} for
short, guarantees that $\cal R$ always receive the sent message and
the adversary does not learn anything about it. It was shown that
PSMT is possible if and only if $n \geq 2t + 1$. See
\cite{DDWY93,SA96,SNR04,ACH06,FFGV07,KS08} for more references.
Franklin and Wright \cite{FW00} relaxed the security requirement of
SMT protocols and proposed \emph{probabilistic} security in which
two parameters $\varepsilon$ and $\delta$ upper bound the advantage
of the adversary in breaking privacy, and the probability that
$\mathcal{R}$ fails to recover the sent message, respectively. In a
PSMT protocol $\varepsilon=\delta=0$. In this paper we refer to
these protocols as \emph{almost SMT} protocols. We refer interested
readers to \cite{DW02,KS07,A08,PCSR08}.

Franklin and Wright \cite{FW00} also considered a model where an
additional reliable broadcast channel is available to $\cal S$ and
$\cal R$.  A broadcast channel guarantees that \emph{all} nodes of
the network receive the same message. We refer to this model as
\emph{Broadcast Model} (BM). They  showed that PSMT in
this model  requires $n \geq 2t + 1$, but probabilistic security can
be obtained with $n> t$ and gave a 3-round $(0, \delta)$ protocol in
this model.

Garay and Ostrovsky \cite{GO08} replaced the broadcast channel with
an authentic and reliable \emph{public channel} that connects $\cal
S$ and $\cal R$. A public channel is totally susceptible to
eavesdropping but is immune to tampering. We refer to this
communication model as \emph{Public Discussion Model} (PDM). Garay
and Ostrovsky \cite{GO08} gave a 4 round protocol with probabilistic
security when $n > t$, which shows that the connectivity requirement
for PDM is the same as the broadcast model.

Efficiency parameters of SMT protocols are, (i) the number of {\em
rounds} where each round is one message flow between $\cal S$ and
$\cal R$, or vice versa, and (ii) the communication efficiency
measured in terms of \emph{transmission rate}  which is the total
number of bits sent over all wires for a message divided by the
length of the secret.

Round complexity in PDM is measured by a pair  $(r, r')$ where $r$
is the total number of rounds and $r'$ is the number of rounds that
the public channel is invoked ($r \geq r'$).

{\em Related models:} Pubic channel has been used in other contexts
including unconditionally secure key agreement \cite{M93} where the
public channel is used for the advantage distillation, information
reconciliation and privacy amplification. The public channel in this
case is  a free resource and its communication cost is not
considered. In PDM however, the cost of realizing a channel in a
distributed system is taken into account.

\newcommand{\tabincell}[2]{\begin{tabular}{@{}#1@{}}#2\end{tabular}}
\begin{table*}
\renewcommand{\arraystretch}{1.0}
\renewcommand{\multirowsetup}{\centering}
\centering \caption{Main results on lower bounds of connectivity and
round of SMT protocols in PDM} \label{tbl:compCons}
\begin{tabular}{|c|c|c|c|c|c|c|}
\hline \textbf{Type} & \tabincell{c}{ \textbf{Resiliency}} &
\tabincell{c}{\textbf{Round}} &
\textbf{Construction} & \tabincell{c}{\textbf{Transmission Rate}}\\
\hline
\tabincell{c}{$(\varepsilon, \delta)$ \\
$\varepsilon+ \delta < 1-\frac{1}{|\mathbf{M}|}$}
& $n \leq 2t$ & (2,2) & \tabincell{c}{Impossible  \\ (\textbf{Theorem} \ref{the:ed1})} & \quad \\
\hline \tabincell{c}{$(\varepsilon, \delta)^*$ \\ $\varepsilon +
\delta < 1 - \frac{1}{|\mathbf{M}|} $\\ and $\delta <\frac{1}{2}(1 -
\frac{1}{|\mathbf{M}|})$} & $n \leq 2t$ & $(r,1), r \geq 3$ &
\tabincell{c}{Impossible \\
(\textbf{Theorem} \ref{the:ed2})} & \\
\hline \tabincell{c}{$(\varepsilon, \delta)$-PD-adaptive$^{**}$ \\
$3\varepsilon + 2\delta < 1 - \frac{3}{|\mathbf{M}|}$} & $n \leq 2t$
& $(3,1)$ &
\tabincell{c}{Impossible \\
(\textbf{Theorem} \ref{the:31imp})} & \\
\hline $(0, \delta)$ & $n > t$ & $(3,2)$ & \tabincell{c}{$\surd$
\\ \cite{FW00,G08},
(\textbf{Theorem} \ref{the:uhf})} & \tabincell{l}{\cite{FW00,G08}:
$O(n)$ on wires and public channel\\\textbf{ours}: $O(n)$ on wires
and $O(1)$ on public channel \\\hspace{0.6cm} when  the length of
message
is $\Omega((n\log \delta)^2)$}\\
\hline
\end{tabular}

\flushleft{$^*$ the invoker of public channel is fixed initially
in the protocol\\
$^{**}$ the invoker of public channel is not fixed initially but
adaptive to real execution of the protocol}
\end{table*}

\subsection{Our Results}
Garay et al. \cite{GO08} proposed a $(4,3)$-round protocol  and
subsequently improved its round complexity to $(3,2)$-round
\cite{G08}. However it was not known if this round complexity was
optimal.

The main result of this paper is to prove that the minimum
values of $r$ and $r'$ for which an $(r,r')$-round
$(\epsilon,\delta)$ protocol can exist are 3 and 2, respectively.
This answers the question of round optimality of almost SMT
protocols in PDM that was raised in \cite{GO08}.

Our results on round optimality are obtained in three steps. We
first prove that there is no $(2,2)$-round $(\varepsilon, \delta)$
protocol in PDM with $\varepsilon+ \delta < 1-{1/|\mathbf{M}|}$ when
$n\leq2t$, where $\mathbf{M}$ denotes the message space. This means
that message transmission protocols in PDM with $(2,2)$-round
complexity will be either unreliable, or insecure.

In the second step we will show that when the invocation of the
public channel does not depend on the protocol execution and is
statically determined as part of protocol description, there is no
$(r \geq 3,1)$-round $(\varepsilon, \delta)$ protocol with
$\varepsilon+ \delta < 1-{1/|\mathbf{M}|}$ and $\delta <
\frac{1}{2}(1-{1/|\mathbf{M}|})$ when $n\leq2t$.

Then we  generalize this result to the case that the invoker of the
public channel is not fixed at the start of the protocol and is
adaptively determined in each execution, and show that there is no
$(3,1)$-round $(\varepsilon, \delta)$ protocol with $3\varepsilon +
2\delta < 1 - {3}/{|\mathbf{M}|}$.

We also construct a round optimal protocol that has constant
transmission rate over the public channel when the length of message
(i.e., $\log|\mathbf{M}|$) is $\Omega((n\log \delta)^2)$ bits long.

Table \ref{tbl:compCons} summarizes our results and puts them in
relation to others' works.

\subsection{Discussion}
One of the main motivations for studying SMT  is to reduce
connectivity requirement in secure multiparty protocols
\cite{BGW88,CCD88,RB89}. Secure multiparty  protocols  require a
secure and reliable channel between every two nodes and so require
the network graph to be \emph{complete}. Using an SMT protocol one
can simulate secure connection between any two nodes using a network
with sufficient connectivity, that is $n$ disjoint paths (and not
direct link) between any two nodes where $n>2t$. Secure message
transmission in PDM  can further reduce connectivity $(n > t)$ as
long as there is an authentic public channel. This is the lowest
possible connectivity and shows that two nodes  can securely
communicate as long as there is one uncorrupted path between them
(and a public channel). Realizing a public channel in an
point-to-point sparse network however is costly. For example it is
possible to simulate such a channel using \emph{almost-everywhere
broadcast protocol} \cite{GO08} that uses \emph{almost-everywhere
Byzantine agreement protocol} \cite{DPPU88}. It is shown
\cite{Upf92} that in degree-bounded networks  agreement on a single
bit using almost-everywhere agreement protocol requires at least
$O(\log N)$ rounds communication, where $N$ is the number of nodes
in the network.

The high cost of simulating the public channel is the motivation for
reducing the number of invocation and transmission rate of such a
channel.

\subsection{Organization}
Section 2 describes the security model and relevant definitions.
Lower bounds on round complexity of SMT protocol in PDM are proved
in Section 3. Section 4 describes an round optimal $(0, \delta)$-SMT
by public discussion protocol. Finally we draw a conclusion in
Section 5.
\section{Preliminaries}
\subsection{Model and Notations}
\noindent \emph{\underline{Network model.}} We assume a
\emph{synchronous}, connected point-to-point \emph{incomplete}
network. Players $\mathcal{S}$ and $\mathcal{R}$ are connected by
$n$ vertex-disjoint paths, called {\em wires}. In addition to the
wires, we assume there is an authentic and reliable \emph{public
channel} between $\mathcal{S}$ and $\mathcal{R}$. Messages over this
channel are publicly accessible and are correctly delivered to the
recipient. All wires and the public channel are bidirectional. SMT
protocols proceed in {\em rounds}. In each round, one player may
send a message on each wire and the public channel, while the other
player will only receive the sent messages. The sent messages will
be delivered before the next round starts. \\
\noindent \emph{\underline{Adversary model.}} The adversary
$\mathcal{A}$ is computationally \emph{unbounded}. $\mathcal{A}$ can
corrupt nodes on paths between $\cal S$ and $\cal R$. A wire is
corrupted if at least one node on the path is corrupted. We assume
up to $t \leq n-1$ wires can be corrupted by the adversary.
$\mathcal{A}$ can \emph{eavesdrop}, \emph{modify} or \emph{block}
messages sent over the corrupted wires. $\mathcal{A}$ is assumed to
be \emph{adaptive}, meaning that she can corrupt wires during the
protocol execution based on the communication traffic it has seen so
far.

We also consider \emph{static} adversary by which we mean that the
adversary chooses the  corrupted wires before the start of the
protocol. A static adversary will however act adaptively during the
protocol execution with regard to messages that are sent over the
corrupted wires: in each round the adversary sees the traffic over
all the corrupted wires and the public channel before tampering the
traffic over the corrupted wires in that round.
\\
\noindent \emph{\underline{Notations.}} Let $\mathbf{M}$ be the
message space. Let $M_S$ denote the secret message of $\cal S$, and
$M_R$ the  message output by $\cal R$. We use $\bot$ to denote null
string and $\emptyset$ to denote empty set. The notation $u
\leftarrow \cal{U}$ denotes that a value $u$ is sampled uniformly
from a set $\cal{U}$.
\subsection{Definitions}
The \emph{statistical distance} of two random variables $X, Y$ over
a set $\cal U$ is given by, \begin{equation}{\rm \Delta}(X, Y) =
\frac{1}{2}\sum_{u \in \cal{U}} \mbox{\Big |}\Pr[X = u] -
\Pr[Y=u]\mbox{\Big |}.\end{equation}

\begin{lemma}{\rm \cite{W06}} \label{lem:sdci}
Let $X,Y$ be two random variables over a set $\cal U$. The advantage
of any computationally unbounded  algorithm %\footnotemark[1]
$\mathcal{D}: \mathcal{U} \to \{0, 1\}$ to distinguish $X$ from $Y$
is $$|\Pr[{\cal D}(X)=1]-\Pr[{\cal D}(Y)=1]|
\leq {\rm \Delta}(X, Y).$$
\end{lemma}
\vspace{0.1cm}

In an \emph{execution} of an SMT protocol ${\rm \Pi}$, $\cal{S}$
wants to send $M_S \in \mathbf{M}$ to $\cal{R}$ privately and
reliably. We assume that at the end of the protocol, $\cal{R}$
\emph{always} outputs a message $M_R \in \mathbf{M}$.

An execution is completely determined by the random coins of all the
players including the adversary, and the message distribution of
$M_S$. For $P \in \{\cal S, R, A\}$, the view of $P$ includes the
random coins of $P$ and the messages that $P$ receives. Denote by
$V_A(m, c_A)$ the view of $\cal A$ when the protocol is run with
$M_S = m$ and $\cal A$'s randomness $C_A=c_A$.
\begin{definition} \label{def:smt}
A protocol  between $\mathcal{S}$ and $\mathcal{R}$ is an
$(\varepsilon, \delta)$-\textsf{{Secure Message Transmission by
Public Discussion}} {\rm (SMT-PD)} protocol if the following two
conditions are satisfied:
\begin{itemize}
  \item {\textbf{Privacy:}}
  For every two messages $m_0, m_1 \in \mathbf{M}$ and $c_A \in \{0, 1\}^*$, it has
  \[{\rm \Delta}(V_A(m_0, c_A), V_A(m_1, c_A)) \leq \varepsilon,\]
  where the probability is taken over the randomness of $\cal{S}$
  and $\cal{R}$.

  \item {\textbf{Reliability:}} $\cal{R}$ recovers the message
  $M_S$ with probability larger than $1- \delta$, or formally \[\Pr[M_R \ne M_S] \leq \delta,\]
  where the probability is over the randomness of
  players $\cal S, R$ and $\cal A$, and the choice of $M_S$.
\end{itemize}
\end{definition}

Observe that the above definition is oblivious of the message
distribution, meaning that given an SMT-PD protocol, it will be
secure with the same privacy and reliability parameters regardless
of the concrete distribution over $\mathbf{M}$.

\section{Round Complexity of SMT-PD Protocol}
By the similarity of broadcast model and public discussion model, we
recall Franklin and Wright's results \cite{FW00} in our language as
follows.
\begin{theorem}{\rm \cite{FW00}} \label{the:fw00}
If $n \leq 2t$, then: {\rm (i)} For any values $r \geq r' $, it is
impossible to construct $(r, r')$-round $(0,0)$-{\rm SMT-PD}
protocols; {\rm (ii)} For any values $r >0$ and $0 \leq \epsilon
\leq 1$, it is impossible to construct $(r, 0)$-round $(\epsilon,
\delta)$-{\rm SMT-PD} protocols with $\delta <
\frac{1}{2}(1-\frac{1}{|\mathbf{M}|})$.
\end{theorem}

In this section, we will prove when $n\leq2t$ any $(\varepsilon,
\delta)$-SMT-PD protocol needs $(3, 2)$-round complexity. This is by
proving that: (i) secure $(2,2)$-round $(\varepsilon,
\delta)$-SMT-PD protocols do \emph{not} exist, and (ii) for any $(3,
1)$-round protocol, either privacy or reliability can be
compromised.

The following lemma plays a central role in proving the
impossibility results in this paper. Loosely speaking, the lemma
shows that for an $(\varepsilon, \delta)$-SMT-PD protocol no
algorithm that is given the adversary's view as the input, can
output $M_S$ with a probability much better than random guess.

\begin{lemma}\label{lem:edide}
Let $\rm \Pi$ be an $(\varepsilon, \delta)$-{\rm SMT-PD} protocol
and  assume $\cal S$ selects $M_S \from \mathbf{M}$. Then no
adversary $\mathcal{A}$ can correctly guess $M_S$ with probability
larger than $\varepsilon + {1/|\mathbf{M}|}$. That is,
\[
\Pr[M_A = M_S] \leq \varepsilon +
{1/|\mathbf{M}|},
\]
where  $M_A$ denotes the adversary's output,  and the probability is
taken over the random coins of $\mathcal{S}, \mathcal{R}$ and $\cal
A$.
\end{lemma}
In proving Lemma \ref{lem:edide}, we need the Lemma \ref{cor:advdis}
below (See Appendix \ref{cor:advdisprf} for its proof).

\begin{lemma} \label{cor:advdis}
Consider an {\rm $(\varepsilon, \delta)$-SMT-PD} protocol ${\rm
\Pi}$ and an  adversary $\cal B$ that plays the following  game: the
challenger $\cal C$ sets up the system;  $\cal B$ selects two
messages $M_0, M_1$ from $\mathbf{M}$ and gives them to a challenger
$\cal C$ who selects $b \from \{0,1\}$ and runs the protocol (by
simulating $\cal S, R$) to transmit $M_b$. $\cal B$ can corrupt up
to $t$ wires and finally outputs a bit $b'$.

Let ${\cal B}^{{\rm \Pi}(M_b)}()$ be the output of $\cal B$ when $b$
is selected by $\cal C$ in the simulation. Then
\begin{equation} \label{eq:advdis}
\left|\Pr[\mathcal{B}^{{\rm \Pi}(M_0)}() = 1] -
\Pr[\mathcal{B}^{{\rm \Pi}(M_1)}()=1]\right| \leq \varepsilon,
\end{equation}
where the probability is taken over the randomness of \  $\cal C$
and $\cal B$.
\end{lemma}
\begin{proof}(of Lemma \ref{lem:edide}) \quad The proof is by
contradiction: assume that there is an adversary $\mathcal{A}$ that
can output $M_A$ with probability $\Pr[M_A = M_S]
> \varepsilon + {1/|\mathbf{M}|}$. We will construct an algorithm
${\cal B}$ to invalidate Eq.(\ref{eq:advdis}) .

The code of ${\cal B}$ is as follows: ${\cal B}$ \emph{randomly}
chooses two messages $(M_0, M_1) \in \mathbf{M}$ and asks its
challenger $\cal C$ to transmit one of the two messages. $\cal C$
chooses a bit $b \from \{0, 1\}$  and simulates $\cal S, R$ to run
protocol $\rm \Pi$ in transmitting $M_b$.   ${\cal B}$ runs
adversary $\cal A$ as a subroutine to attack the protocol. ${\cal
B}$  answers $\cal A$'s queries  by
forwarding them to the challenger and returning the
results back to $\mathcal{A}$. At the end of the
protocol $\mathcal{A}$ outputs a message in $\mathbf{M}$ (which can
be different from $M_1$ and $M_0$). ${\cal B}$ outputs 1 if
$\mathcal{A}$ outputs $M_1$, and outputs 0, otherwise. Note that
${\cal B}$ will have the complete view of $\cal A$. Then
$$
\begin{array}{lll}
&&\hspace{-1.1cm}\Pr[{\cal B}^{{\rm \Pi}(M_1)}()=1] \\
&=& \Pr[M_A = M_1  \mid {\cal C} \mbox{ has chosen } M_1 ] >
\varepsilon + {1/|\mathbf{M}|}, \end{array}
$$
and
\begin{equation} \label{eq:advreduc2}
\begin{array}{lll}
&&\hspace{-1.3cm}\Pr[{\cal B}^{{\rm \Pi}(M_0)}()=1] \\
&=& \Pr[M_A = M_1 \mid {\cal C} \mbox{ has chosen } M_0 ] =
{1/|\mathbf{M}|}.
\end{array}
\end{equation}

Note that Eq.(\ref{eq:advreduc2}) follows by that fact that $M_1$ is
chosen independent of $M_0$ and the randomness of players $\cal S$
and $\cal R$ in the simulation of $\cal C$ and so the probability of
$\cal A$'s output to be equal to $M_1$ (which is chosen randomly) is
at most the probability of random guess which is $
{1/|\mathbf{M}|}$. Hence, we have $\Pr[{\cal B}^{{\rm
\Pi}(M_1)}()=1] - \Pr[{\cal B}^{{\rm \Pi}(M_0)}()=1] > \varepsilon,$
contradicting Corollary \ref{cor:advdis}.
% Thus
%completes the proof.
%$\hfill\blacksquare$
\end{proof}

\subsection{Impossibility of $(2,2)$-Round $(\varepsilon, \delta)$-SMT-PD Protocol when $n \leq 2t$}
The impossibility proof needs to analyze the actions of the
adversary in rounds, hence we start by decomposing an SMT-PD
protocol into rounds as follows.
\begin{definition} \label{def:edfunc}
For a $(r, r')$-round {\rm SMT-PD} protocol, the functionality of
the protocol is described as a sequence of randomized functions
$(f_1, \ldots, f_r, g)$.

The function $f_i$ denotes the {\rm \textsf{round encoding
function}} that is used to generate the traffic sent in the $i$-th
round. The input of $f_i$ consists of the received messages of
previous rounds and random coins of the caller. For a player $P
\in\{\cal S, R\}$, $C_P$ denotes the random coins of $P$, and
$\mathsf{M}_P^i$ denotes the set of all messages received by $P$
during the first $i$ rounds with $\mathsf{M}_S^0 = \{M_S\}$ and
$\mathsf{M}_R^0 = \emptyset$. If the initiator of round $1 \leq i
\leq r$ is $P$, we write $P_iX_iY_i = f_i(\mathsf{M}_P^{i-1}, C_P)$
to denote the random variable corresponding to traffic in round $i$;
here $P_i$ denotes the traffic over the public channel, and $X_i$
and $Y_i$ denote the traffic over the corrupted wires and the
uncorrupted wires, respectively, or vice versa.

The function $g$ denotes the {\rm \textsf{decoding function}}. By
the end of the protocol $\cal R$ outputs $M_R = g(\mathsf{M}_R^r,
C_R)$.
\end{definition}
\vspace{0.1cm}

\begin{theorem}\label{the:ed1}
Let $n \leq 2t$. Then there is no $(2,2)$-round $(\varepsilon,
\delta)$-{\rm SMT-PD} protocol with $\varepsilon+ \delta <
1-{1/|\mathbf{M}|}$.
\end{theorem}

The proof is by contradiction: suppose there exists a $(2,2)$-round
$(\varepsilon, \delta)$-SMT-PD protocol $\rm \Pi$ with $\varepsilon+
\delta < 1-{1/|\mathbf{M}|}$. We construct an adversary $\cal A$
that breaks the \emph{privacy} of $\rm \Pi$ by impersonating $\cal
R$. We show that for each execution of $\rm \Pi$ where $\cal S$
sends a message $m$ to $\cal R$, there exists a second execution
called {\em swapped execution} where $\cal S$  sends the message $m$
but  $\cal A$ impersonates $\cal R$  such that $\cal S$ receives
identical traffic in the two executions and so cannot distinguish
the two. The views of $\cal R$ and $\cal A$ are however swapped in
the two executions, and so if $\cal R$ outputs $M_R = M_S$ in one of
the executions, then $\cal A$ outputs $M_A = M_S$ in the swapped
execution and so $\Pr[M_A = M_S] \geq \Pr[M_R = M_S]$. Using Lemma
\ref{lem:edide} and that $\rm \Pi$ is an $(\epsilon,\delta)$-SMT-PD
protocol, we have $\varepsilon+ \delta \geq 1-{1/|\mathbf{M}|}$
which is a contradiction.
\begin{figure*}[!ht]
$$
\input xy
\xyoption{all}  \xymatrix@R=0.05in@C=1.1in {
\mathcal{\underline{S}}_{\quad(M_S, C_S)}&
\mathcal{\underline{A}}_{\quad \left( C_{A0}, C_{A1} \right)}&
\mathcal{\underline{R}}_{\quad(C_R)} \\
\txt<8pc>{} &\ar[l]_-{\large P_1X_1Y_1'} \txt<8pc>{finds $C_R'$,
$P_1X_1'Y_1'=f_1(C_R')$}
&\ar[l]_-{\large P_1X_1Y_1} \txt<7pc>{$P_1X_1Y_1=f_1(C_R)$} \\
\txt<7pc>{$P_2X_2Y_2=f_2(M_S, P_1X_1Y_1', C_S)$}
\ar[r]^-{P_2X_2Y_2}&*\txt<8pc>{blocks $Y_2$, computes\\
$M_{A}=g(P_2Y_2, C_R')$}
\ar[r]^-{P_2X_2}&*\txt<8pc>{$M_R = g(P_2X_2, C_R)$} \\
}
$$
\caption{An execution $E$ of $\rm \Pi$ in the presence of adversary
$\cal A$ with $C_{A0}=1$.}\label{fig:exeE}
$$
\input xy
\xyoption{all}  \xymatrix@R=0.05in@C=1.1in {
\mathcal{\underline{S}}_{\quad(M_S, C_S)}&
\mathcal{\underline{A}}_{\quad\left(C_{\hat{A}0}, C_{\hat{A}1}
\right)}&
\mathcal{\underline{R}}_{\quad(C_{\hat{R}})} \\
\txt<8pc>{} &\ar[l]_-{\large P_1X_1Y_1'}
\txt<8pc>{$C_{\hat{R}}'=C_R$, $P_1X_1Y_1=f_1(C_{\hat{R}}')$}
&\ar[l]_-{\large P_1X_1'Y_1'} \txt<7pc>{$C_{\hat{R}}=C_R'$, $P_1X_1'Y_1'=f_1(C_{\hat{R}})$} \\
\txt<7pc>{$P_2X_2Y_2=f_2(M_S,P_1X_1Y_1',C_S)$}
\ar[r]^-{P_2X_2Y_2} &*\txt<8pc>{blocks $X_2$, computes \\
$M_{\hat{A}}=g(P_2X_2, C_{\hat{R}}')$}
\ar[r]^-{P_2Y_2}&*\txt<8pc>{$M_{\hat{R}}=g(P_2Y_2, C_{\hat{R}})$} \\
}
$$
\caption{The swapped execution $\hat{E}$ of $E$ with
$C_{\hat{A}0}=0$ and $C_{\hat{R}}=C_R', C_{\hat{R}}' =
C_R$.}\label{fig:exehatE}
\end{figure*}
\begin{proof}
Assume by contradiction that there is a $(2,2)$-round $(\varepsilon,
\delta)$-SMT-PD protocol ${\rm {\Pi}}$ with $\varepsilon+ \delta <
1-{1/|\mathbf{M}|}$, and the message distribution over $\mathbf{M}$
is uniform. Suppose wires are labeled by $1,2,\dots, n$, and $n=2t$.
(Note if there exists an $(\varepsilon, \delta)$-SMT-PD protocol for
$n'<2t$, the same protocol can be run for $n=2t$ by neglecting the
last $n-n'$ wires. Thus an impossibility result for $n=2t$ still
holds for $n'<2t$.)

The adversary is assumed to be \emph{static} in the following. That
is, the corrupted wires are selected at the start of the protocol.
The impossibility results obtained for such adversary will hold for
more powerful \emph{adaptive} adversaries who will corrupt the wires
during the running of the protocol.

We write $\cal A$'s randomness as $C_A = (C_{A0}, C_{A1})$  where
$C_{A0} \in \{0, 1\}$ is used to select one of the two  sets of $t$
wires: $\{1,\dots, t\}$ or $\{ t+1, \dots, 2t\}$ for corruption and
$C_{A1} \in \{0, 1\}^*$ is used for  encoding and decoding of the
traffic. Let $C_{A0}=0$ and $C_{A0}=1$ denote  the first and the
last $t$ sets of wires will be corrupted, respectively.

Before going ahead, we remark that: (i) The last round message of a
SMT-PD protocol can only be from $\mathcal{S}$ to $\cal R$ as
otherwise it can be removed without affecting the output of
$\mathcal{R}$. (ii) For generality we don't assume the interaction
in a SMT-PD protocol should be back-and-forth, meaning that some
consecutive rounds of the protocol may have the same sender and
cannot be combined into one round. Under the effect of public
channel, this provides a possible paradigm in designing SMT-PD
protocols. E.g., both of the first two rounds of the protocol in
\cite{GO08} are from $\cal S$ to $\cal R$, and are from $\cal R$ to
$\cal S$ in \cite{G08}.

Therefore, depending on the order of the first round, a 2-round
SMT-PD protocol has two kinds of interactions.
\\[0.1cm]
\underline{CASE 1}. In this case, the first round traffic is from
$\cal R$ to $\cal S$, while the second round is from $\cal S$ to
$\cal R$. Assume $C_{A0}=1$, i.e., the last $t$ wires are corrupted.
We illustrate the strategy of $\cal A$ in Fig. \ref{fig:exeE} and
formalize it as follows.
\begin{itemize}
\item{Round 1:}
When $\mathcal{R}$ sends $P_1X_1Y_1 = f_1(C_R)$; $\mathcal{A}$
computes $P_1X_1'Y_1' = f_1(C_R')$ where $C_R'$ is the value
computed from $C_{A1}$  and results in  $P_1$ over the public
channel, hence $\mathcal{A}$ can leave the transmission over the
public channel unchanged. This is always possible because the
function table of $f_1$ is public and $\cal A$ is computationally
unbounded. Thus $\cal A$ can find the set of random strings such
that $\Omega = \{r \mid f_1(r) = P_1X_1'Y_1'\}$ and selects $C_R'
\from \Omega$. $\cal A$ will then replaces $Y_1$ by $Y_1'$.

\item{Round 2:} When $\cal{S}$ generates message $P_2X_2Y_2 =
f_2(M_S, P_1X_1Y_1', C_S)$, $\cal{A}$ blocks the transmission over
the corrupted wires and outputs $M_A = g(P_2Y_2, C_R')$.
\end{itemize}

Let $\mathbf{E}$ be the set of all executions of ${\rm {\Pi}}$ in
presence of $\mathcal{A}$. We consider a binary relation
$\mathbf{W}$ over $\mathbf{E}$ such that $(E, \hat{E}) \in
\mathbf{W}$ if, (i) $M_S, C_S$ are the same in the two executions;
(ii) $C_{\hat{A}0} \oplus C_{A0} = 1$; and (iii) $C_{\hat{R}} =
C_R', C_{\hat{R}}' = C_R$, where ` $\hat{}$ ' in the superscript
denotes the random coins used and messages output by $\cal A$ and
$\cal R$ in $\hat{E}$, respectively. Note that in the two
executions, the $t$ corrupted wires are swapped with the uncorrupted
ones such that the messages received by $\cal A$ and $\cal R$ are
swapped as shown in Fig. \ref{fig:exeE} and \ref{fig:exehatE}.

For a pair of $(E, \hat{E}) \in \mathbf{W}$, the first round
messages received by $\cal S$ in $E$ and $\hat{E}$ are identical and
equal to $P_1X_1Y_1$. Thus in the second round, $\cal S$ will
generate the same traffic $P_2X_2Y_2$ in both $E$ and $\hat{E}$, and
so if $\cal R$ outputs $M_R$ in $E$, $\cal A$ will output
$M_{\hat{A}} = M_R$ in $\hat{E}$ since $M_R = g(P_2X_2, C_R) =
g(P_2X_2, C_{\hat{R}}') = M_{\hat{A}}$.

Let $p_E$ be the probability that execution $E$ is running.
Similarly define $p_{\hat{E}}$. Denote by $\mathbf{S} \subseteq
\mathbf{E}$ the set of executions with $M_R = M_S$ and so we have
$\Pr[M_R = M_S] = \sum_{E \in \mathbf{S}}p_{E}.$ Now $M_{\hat{A}} =
M_S$ holds in $\hat{E}$ if $M_R = M_S$ holds in $E$ and so we have
$\Pr[M_A = M_S] \geq \sum_{E \in \mathbf{S}} p_{\hat{E}}.$

Observe that $p_E$ is completely determined by the probability of
selecting $M_S$ and other random coins of all the players. For any
two executions $(E, \hat{E}) \in \mathbf{W}$, we note that $(M_S,
C_S) = (M_{\hat{S}}, C_{\hat{S}})$, while $C_R$ and $C_{\hat{R}}$
are both selected with uniform probability. Moreover, when $C_R$ and
$C_{\hat{R}}$ are fixed, both of the probability of selecting $C_A$
and $C_{\hat{A}}$ are $2^{-1-\lceil \log |\Omega| \rceil}$. We thus
get $p_E = p_{\hat{E}}$.

Then by Lemma \ref{lem:edide} and above argument,
\begin{eqnarray} \label{eq:mrmsmams}
1-\delta \leq \Pr[M_R = M_S]  \leq \Pr[M_A = M_S]\leq 1/|\mathbf{M}|
+\epsilon .
\end{eqnarray}
Therefore, it has $\varepsilon + \delta
\geq 1 - {1/|\mathbf{M}|},$ which contradicts the assumption on $\rm
\Pi$. \vspace{0.2cm}

\noindent \underline{CASE 2}. In this case, both of the two rounds
traffic are from $\cal S$ to $\cal R$. Intuitively, if $n \leq 2t$
and $\cal S$ receives no feedback from $\cal R$, $\cal A$ can just
block the traffic over the $t$ corrupted wires such that $\cal R$
has no advantage over $\cal A$ in recovering $M_S$.

More specifically, considering two executions $E$ and $\hat{E}$ in
this case, where the random coins of $\cal A$ and $\cal R$ are
swapped, and the corrupted and uncorrupted wires are also swapped.
If $\cal A$ blocks the $t$ corrupted wires, the view of $\cal R$ in
$E$ will equal the view of $\cal A$ in $\hat{E}$. Then if $\cal R$
outputs $M_S$ in one execution, $\cal A$ will output it in the
swapped execution. By Lemma \ref{lem:edide} and the assumption on
$\rm \Pi$, Eq. (\ref{eq:mrmsmams}) holds also in this case, thus it
follows that $\varepsilon + \delta \geq 1 - {1/|\mathbf{M}|}$.
\end{proof}
\subsection{Impossibility of $(r, 1)$-Round $(\varepsilon, \delta)$-SMT-PD Protocol when $n \leq 2t$}
Theorem \ref{the:ed1} shows that optimal $(\epsilon, \delta)$-SMT-PD
protocols need at least 3 rounds, while Theorem \ref{the:fw00} shows
that at least one round public channel invocation is necessary. A
natural question thus is to find out if secure $(r \geq 3, 1)$-round
SMT-PD protocols can exist. As a warm-up, the following theorem
gives a negative answer to the case that the invoker of public
channel is specified initially in the protocol.

\begin{theorem}\label{the:ed2}
Let $n \leq 2t$ and $r \geq 3$. Then a $(r, 1)$-round $(\varepsilon,
\delta)$-{\rm SMT-PD} protocol with fixed invoker of public channel
has either $\varepsilon + \delta \geq 1 - \frac{1}{|\mathbf{M}|} $
or $\delta \geq \frac{1}{2}(1 - \frac{1}{|\mathbf{M}|})$.
\end{theorem}

The proof is by contradiction: assume there exists a $(r,1)$-round
$(\varepsilon, \delta)$-{\rm SMT-PD} protocol $\rm \Pi$ with fixed
public channel invoker, where values of $\varepsilon$ and $\delta$
do not satisfy any of the above inequalities. We construct an
adversary who can break either the privacy or the reliability of
$\rm \Pi$.

$\cal A$'s strategy is to block the traffic (over the $t$ corrupted
channels) sent by the invoker of public channel, and to replace the
traffic (over the $t$ corrupted wires) sent to the invoker by forged
traffic that is constructed according to the protocol description.
Then,
\begin{enumerate}
\item If the public channel is invoked by  $\cal S$, we will show that $\cal S$ cannot
distinguish two swapped executions in which she has the same views.
The two executions have the property that if $\cal R$ outputs $M_R =
M_S$ in one execution then $\cal A$ outputs $M_A = M_S$ in the
swapped execution. Using an argument similar to Theorem
\ref{the:ed1} we prove that the adversary can break the
\emph{privacy} of the protocol and thus obtain $\varepsilon + \delta
\geq 1 - \frac{1}{|\mathbf{M}|} $.

\item If the public channel is invoked by  $\cal R$, we will show that $\cal R$ cannot distinguish
two swapped executions in which he has the same views. If in one
execution $\cal R$ outputs $M_S$, he will output $M_A$ in the
swapped execution with the same probability. The two executions have
the same probability and so when $M_S \ne M_A$, we prove the
adversary can break the \emph{reliability} of the protocol and so
obtain $\delta \geq \frac{1}{2}(1 - \frac{1}{|\mathbf{M}|})$.
\end{enumerate}
\begin{proof}
We stress that in this proof the invoker of the public channel is
already specified in the protocol, whereas the actual invocation
round of the public channel can be adaptive to the protocol
execution. The impossibility result will hold straightforwardly for
the case that the invocation round of the public channel is a part
of the protocol specification.

As noted in the proof of Theorem \ref{the:ed1}, the interaction
order in the protocol is not necessarily back-and-forth, and the
last round is from $\cal S$ to $\cal R$. Moreover, we also suppose
the message distribution over $\mathbf{M}$ is uniform, and $n=2t$
and the adversary is \emph{static}.

We separate the randomness $C_A$ (of $\cal A$) into four parts: $
(C_{M_A}, C_{A0}, C_{A1}, C_{A2})$, where $C_{A0} \in \{0, 1\}$ is
used to choose one of the two subsets of $t$ wires to corrupt
($C_{A0}=0$ and $C_{A0}=1$ are used for the first or the last $t$
wires, respectively), $C_{A1}$ is used to generate traffic for
substituting the message sent by $\cal S$, $C_{A2}$ for generating
traffic to substitute the message sent by $\cal R$, and $C_{M_A}$
denotes the randomness of $\cal A$ uniformly selecting a message
from $\mathbf{M}$ to impersonate $\cal S$'s traffic.

\begin{figure*}
$$
\input xy
\xyoption{all}  \xymatrix@R=0.08in@C=0.9in {
\mathcal{\underline{S}}_{\quad(M_S,C_S)}&\mathcal{\underline{A}}_{
\quad(C_{M_A}, C_{A0}, C_{A1}, C_{A2})}&\mathcal{\underline{R}}_{\quad(C_R)} \\
\txt<8pc>{$X_1Y_1=f_1(M_S, C_S)$} \ar[r]^-{X_1Y_1}&*\txt<9pc>{blocks
$Y_1$}
\ar[r]^-{X_1}&*\txt<10pc>{} \\
\txt<7pc>{} &\ar[l]_-{\large X_2Y_2'}
\txt<9pc>{$X_2'Y_2'=f_2(Y_1,C_{A2})$} &\ar[l]_-{\large X_2Y_2}
\txt<10pc>{$X_2Y_2=f_2(X_1,C_R)$}\\
\txt<7pc>{} &\ar[l]_-{\large X_3Y_3'}
\txt<9pc>{$X_3'Y_3'=f_3(Y_1,C_{A2})$} &\ar[l]_-{\large X_3Y_3}
\txt<10pc>{$X_3Y_3=f_3(X_1,C_R)$}\\
\txt<7pc>{\vdots} &*\txt<9pc>{\vdots} &*\txt<9pc>{\vdots}\\
\txt<8pc>{$P_iX_iY_i=f_i(M_S,X_2Y_2',\dots, C_S)$}
\ar[r]^-{P_iX_iY_i}&*\txt<9pc>{blocks $Y_i$} \ar[r]^-{P_iX_i}&*\txt<10pc>{} \\
\txt<7pc>{} &\ar[l]_-{\large X_{i+1}Y_{i+1}'}
\txt<9pc>{$X_{i+1}'Y_{i+1}'=f_{i+1}(Y_1, \dots, C_{A2})$}
&\ar[l]_-{\large X_{i+1}Y_{i+1}}
\txt<10pc>{$X_{i+1}Y_{i+1}=f_{i+1}(X_1, \dots, C_R)$} \\
\txt<8pc>{$X_{i+2}Y_{i+2}=f_{i+2}(M_S, X_2Y_2', \dots, C_S)$}
\ar[r]^-{X_{i+2}Y_{i+2}}&*\txt<9pc>{blocks $Y_{i+2}$}
\ar[r]^-{X_{i+2}}&*\txt<10pc>{} \\
\txt<7pc>{\vdots} &\ar[l]_-{} \txt<9pc>{\vdots} &\ar[l]_-{}
\txt<10pc>{\vdots} \\
\txt<8pc>{$X_rY_r=f_r(M_S, X_2Y_2', \dots, C_S)$}
\ar[r]^-{X_rY_r}&*\txt<10pc>{blocks $Y_r$}
\ar[r]^-{X_r}&*\txt<10pc>{$M_R = g(X_1, \dots, X_r, C_R)$} }
$$
\caption{The behaviors of $\cal A$ in an execution where the public
channel is used by $\cal S$ and $C_{A0} = 1$.}\label{fig:genExe}
\end{figure*}
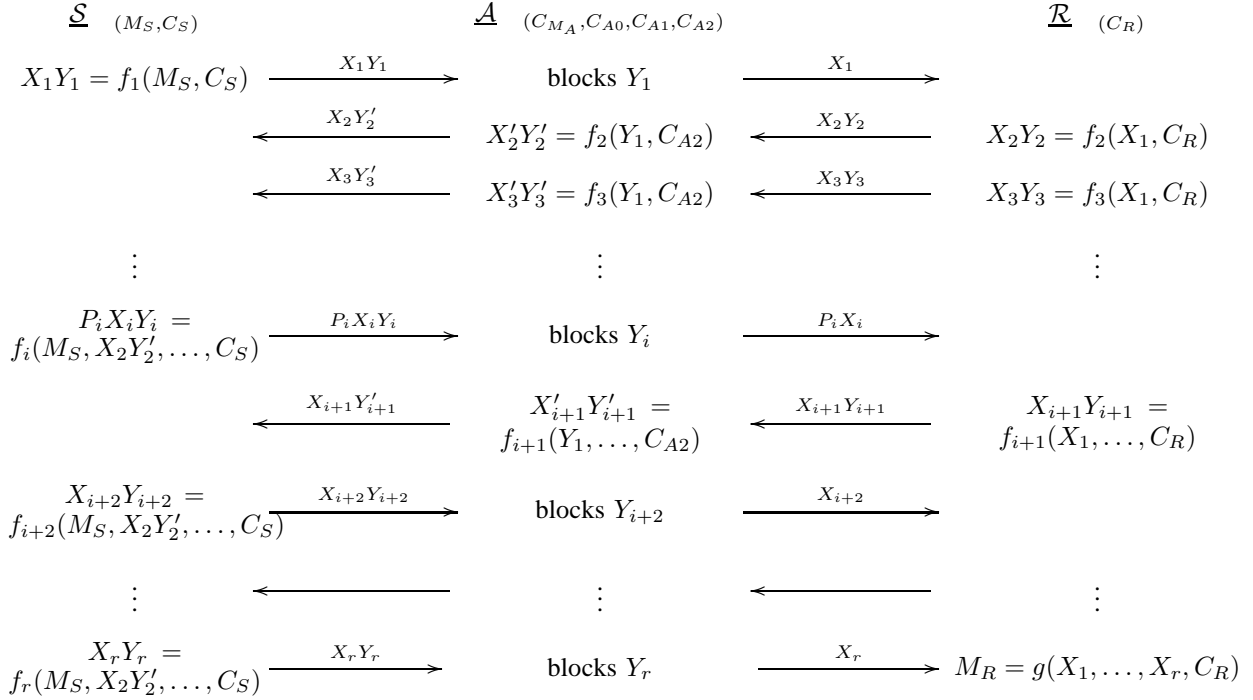

\vspace{0.2cm} \noindent \underline{CASE 1.} [$\cal S$ invokes the
public channel.] \quad We show that in this case $\cal A$ will break
the \emph{privacy} of $\rm \Pi$. Without loss of generality, assume
$C_{A0}=1$. We describe the action of $\cal A$ as follows: in round
$1 \leq j \leq r$,
\begin{itemize}
\item When $\cal S$ sends $X_jY_j$ or $P_jX_jY_j$, $\cal A$ blocks $Y_j$.
\item When $\cal R$ sends $X_jY_j$,
$\cal A$  computes $X_j'Y_j' = f_j(\mathsf{M}_{A}^{j-1}, C_{A2})$,
then replaces $Y_j$ by $Y_j'$. (Here $\mathsf{M}_{A}^{j-1}$ denotes
the messages eavesdropped by $\cal A$ during the first $j-1$
rounds.)
\end{itemize}
Finally, $\cal A$ outputs $M_A=g(\mathsf{M}_{A}^{r},C_{A2})$.

The above strategy of $\cal A$ is also shown in
Fig.\ref{fig:genExe}. Note that $\cal A$ can block and forge
messages as above since $\cal A$ can randomly select $C_A$ to
generate messages $\{X_j'Y_j'\}$, and make them consistent with the
requirement of protocol $\rm \Pi$. Also note that $C_{M_A}=\bot$ and
$C_{A1}=\bot$ since $\cal A$ needs \emph{not} to impersonate $\cal
S$ in this case.

Let $\mathbf{E}$ be the set of  executions of ${\rm {\Pi}}$. We
define a binary relation $\mathbf{W}_1$  over $\mathbf{E}$ to
specify two executions $E$ and $\hat{E}$ as follows: $(E, \hat{E})
\in \mathbf{W}_1$ if: (i) $(M_S, C_S)$ are the same for both
executions; (ii) $C_{\hat{A}0} \oplus C_{A0} = 1$; and (iii) $C_{A2}
= C_{\hat{R}}$ and $ C_R = C_{\hat{A}2}$.
\begin{myclaim}\label{cla:2.1}
{\rm (i)}The view of $\cal S$  in $E$ is the same as her view  in
$\hat{E}$; and {\rm (ii)}the view of $\cal A$ in $\hat{E}$ is
identical to the view of $\cal R$ in $E$. Thus the output of $\cal
R$ in $E$ is the same as the output of $\cal A$ in $\hat{E}$. That
is, $M_R = M_{\hat{A}}$ holds.
\end{myclaim}
\begin{proof}
Without loss of generality assume in execution $E$ we have $C_{A0} =
1$  and the public channel is used in round $i$. Also assume during
the first $i-1$ rounds, $\cal R$ is the initiator of rounds $\{r_1,
\dots, r_\ell\} \subseteq \{1, \ldots, i-1\}$, ordered
nondecreasingly. We first prove statements (i) and (ii) hold during
the first $r_\ell$ rounds, then using the same technique we will
prove the statements hold in the later rounds and thus prove $M_R =
M_{\hat{A}}$.

The proof is by induction over $\ell$. When $\ell=0$, the statements
(i) and (ii) hold trivially from the facts that $\cal S$ doesn't
receive messages in the first $i-1$ rounds and $C_{\hat{A}0} \oplus
C_{A0} = 1$.

For each $j < r$, suppose that the statements (i) and (ii) hold in
the first $r_j$ rounds for $\ell=j$. The induction hypothesis states
that $\mathsf{M}_R^{r_{j}}=\{X_k\}_{k<r_j}$ and
$\mathsf{M}_A^{r_{j}}=\{Y_k\}_{k<r_j}$ are swapped, while
$\mathsf{M}_S^{r_{j}}$ are the same in executions $E$ and $\hat{E}$.
Our objective is to prove that the statements (i) and (ii) also hold
during the first $r_\ell$ rounds for $\ell = j + 1$. Note that in
all those rounds $k$ for $r_j <k <r_{j+1}$, transmissions are only
from $\cal S$ to $\cal R$. Formally the message of each round $k$ is
$X_kY_k = f_k(\mathsf{M}_S^{r_{j}}, C_S)$, and $\cal R$ and $\cal A$
will receive $\{X_k\}_{r_j < k<r_{j+1}}$ and $\{Y_k\}_{r_j <
k<r_{j+1}}$ respectively. Thus $\mathsf{M}_R^{r_{j+1}-1} =
\mathsf{M}_R^{r_{j}} \cup \{X_k\}_{r_j < k<r_{j+1}}$ and
$\mathsf{M}_A^{r_{j+1}-1} = \mathsf{M}_A^{r_{j}} \cup \{Y_k\}_{r_j <
k<r_{j+1}}$. As $C_{\hat{A}0} \oplus C_{A0} = 1$, it follows that
$\mathsf{M}_R^{r_{j+1}-1}$ and $\mathsf{M}_A^{r_{j+1}-1}$ are
swapped in $E$ and $\hat{E}$. Let
$X_{r_{j+1}}Y_{r_{j+1}}'=f_{r_{j+1}}^{(1)}(\mathsf{M}_R^{r_{j+1}-1},
C_R) f_{r_{j+1}}^{(2)}(\mathsf{M}_A^{r_{j+1}-1}, C_{A2})$ be the
messages received by $\cal S$ in round $r_{j+1}$ of $E$. Then $\cal
S$ will receive the same messages in round $r_{j+1}$ of $\hat{E}$
because $C_{A2} = C_{\hat{R}}$, $C_R = C_{\hat{A}2}$, and then
$\mathsf{M}_R^{r_{j+1}-1}$ and $\mathsf{M}_A^{r_{j+1}-1}$ are
exchanged in $E$ and $\hat{E}$. Thus the statements (i) and (ii)
hold during the first $r_{j+1}$ rounds.

Henceforth, $\cal S$ will send $X_k Y_k = f_k(\mathsf{M}_S^k,
C_S)=f_k(\mathsf{M}_S^{r_\ell}, C_S)$ in each later round $k$ for
$r_\ell < k < i$. Observe that in these rounds $\cal S$ won't
receive messages from $\cal R$. Thus if $\cal S$ invokes the public
channel in round $i$ of $E$, it will do the same in $\hat{E}$. And
it follows that the view of $\mathsf{M}_R^i$ and $\mathsf{M}_A^i$ in
$E$ and $\hat{E}$ are swapped during the first $i$ rounds. A similar
argument shows that after the $i$-th round $\cal S$ will receive
identical messages in the two swapped executions. Finally, the views
of $\cal S$ in the two executions will be the same, but
$\mathsf{M}_R^r$ and $\mathsf{M}_A^r$ are swapped in $E$ and
$\hat{E}$. At the end of the protocol, we have $M_R =
g(\mathsf{M}_R^r, C_R) = g(\mathsf{M}_{\hat{A}}^r, C_{\hat{A}2}) =
M_{\hat{A}}$, where $\mathsf{M}_{\hat{A}}^r$ denotes the messages
that $A$ has eavesdropped in execution $\hat{E}$. %$\hfill\Box$
\end{proof}

Let $\mathbf{S}_1 \in \mathbf{E}$ be the set of all successful
executions in which $\cal R$ outputs $M_R = M_S$, and $p_E$ denotes
the  probability of  execution $E$ determined by the random coins of
all players. Define $p_{\hat{E}}$ similarly. Then $\Pr[M_R = M_S] =
\sum_{E \in \mathbf{S}_1}p_{E}.$ By Claim \ref{cla:2.1}, if $E \in
\mathbf{S}_1$, $\cal A$ will output $M_S$ in the swapped execution
of $\hat{E}$; therefore $\Pr[M_A = M_S] \geq \sum_{E \in
\mathbf{S}_1}p_{\hat{E}}.$

Additionally, by the definition of $\mathbf{W}_1$ and the
observation of $C_{M_A} = C_{A1} = \bot$ in this case, we have,
\begin{equation}\label{equ:pephe}
p_E = \frac{1}{|\mathbf{M}|}2^{-r_S-r_R-r_{A2}-1} = p_{\hat{E}},
\end{equation}
where $r_S, r_R, r_{A2}$ denote the length of the random coins  of
$C_S, C_R, C_{A2}$ used by $\cal S, R$ and $\cal A$ respectively.

Now by Eq.(\ref{equ:pephe}), and Lemma \ref{lem:edide}, it follows
that Eq.(\ref{eq:mrmsmams}) also holds in this case, then it yields
that $1 - \frac{1}{|\mathbf{M}|} \leq \varepsilon + \delta$,
contradicting the assumption on $\rm \Pi$.
\\[0.2cm]
\noindent \underline{CASE 2.} [$\cal R$ invokes the public channel.]
\quad We will show that in this case the \emph{reliability} of $\rm
\Pi$ will be broken.  This is by showing that for every successful
execution there exists an unsuccessful one and so probability of
success is at most $1/2$.

Formally, the strategy of $\cal A$ is similar to CASE 1, that is
when $C_{A0}=1$, then in each round $1 \leq j \leq r$:
\begin{itemize}
\item When $\cal R$ sends $X_jY_j$ or $P_jX_jY_j$, $\cal A$ blocks $Y_j$.
\item When $\cal S$ sends $X_jY_j$,
$\cal A$  computes $X_j'Y_j' = f_j(\mathsf{M}_{A}^{j-1}, C_{A1})$
and replaces $Y_j$ by $Y_j'$. (Here $\mathsf{M}_{A}^{j-1}$ denotes
the messages selected and eavesdropped by $\cal A$ during the first
$j-1$ rounds.)
\end{itemize}

Note that $C_{A2}=\bot$ in this case. \emph{For simplicity, we abuse
the notation $M_A$ here to denote the uniformly selected message of
$\cal A$ using coins $C_{M_A}$.}

Let $\mathbf{E}$ and $p_E$ be as defined in CASE 1 and consider a
binary relation $\mathbf{W}_2$ over $\mathbf{E}$  where $(E,
\hat{E}) \in \mathbf{W}_2$ if: (i) $C_R$ is the same in the two
executions; (ii) $C_{\hat{A}0} \oplus C_{A0} = 1$; and (iii) $C_{A1}
= C_{\hat{S}}, C_S = C_{\hat{A}1}$; (iv) $M_S = M_{\hat{A}}$ and $
M_A = M_{\hat{S}}$. Denote by $\mathbf{S}_2$ the set of
\emph{successful} executions in which $\cal{R}$ outputs $M_R = M_S$
under the condition that $M_A \ne M_S$.
\begin{myclaim}\label{cla:2.2}
For each swapped execution pair $(E, \hat{E}) \in \mathbf{W}_2$, the
views of $\cal R$ in $E$ and $\hat{E}$ are identical and so if $E
\in \mathbf{S}_2$ is a successful execution, then $\hat{E} \notin
\mathbf{S}_2$ is a failed execution.
\end{myclaim}
\begin{proof}
Without loss of generality, assume $\cal R$ invokes the public
channel in round $i$ of $E$, and during the first $i$ rounds $\cal
S$ is the initiator of rounds $\{r_1, \ldots, r_\ell\} \subseteq
\{1, \ldots, i-1\}$ (ordered in nondecreasing order) in execution
$E$. By induction on $\ell$, we can prove that $\cal R$ will receive
the same messages during the first $r_\ell$ rounds of the two
swapped executions. This means that $\cal R$ will invoke the public
channel in the same round $i$ of $E$ and $\hat{E}$, both.
Furthermore, we can prove $\cal R$ will receive the same messages
during the later rounds of the two executions. Thus, we have
$\mathsf{M}_{R}^r = \mathsf{M}_{\hat{R}}^r$, where
$\mathsf{M}_{\hat{R}}^r$ denotes all messages that $\cal R$ received
in $\hat{E}$. The  proof is similar to Claim \ref{cla:2.1}.

Now because $M_S$ and $M_{A}$ are swapped in $E$ and $\hat{E}$, if
$\cal{R}$ outputs $M_R = g(\mathsf{M}_R^r, C_R) = M_S$ in $E$, he
will output $M_{\hat{R}} = g(\mathsf{M}_{\hat{R}}^r, C_{R}) =
M_{\hat{A}}=M_S$ in $\hat{E}$. Thus  for any two swapped executions
$(E, \hat{E}) \in \mathbf{W}_2$ when $M_A \ne M_S$, we have $\hat{E}
\notin \mathbf{S}_2$. %$\hfill\Box$
\end{proof}

\begin{myclaim}\label{cla:2.3}
{\rm (i)} The occur probability of any two swapped executions $(E,
\hat{E}) \in \mathbf{W}_2$ is the same; that is $p_E = p_{\hat{E}}$;
and {\rm (ii)} When $M_S \ne M_A$, the failure probability of $\cal
R$ in recovering the secret message is not less than the success
probability of $\cal R$;  formally
\begin{eqnarray*}
&&\hspace{-1.2cm} \Pr[M_R = M_S \mid M_S \ne M_A] \\
&\leq& \Pr[M_R \ne M_S \mid M_S \ne M_A],
\end{eqnarray*}
where the probability is taken over the random coins and messages
selected by $\cal S, R$ and $A$.
\end{myclaim}

\begin{proof}
(i) Note that an execution $E \in \mathbf{E}$ is completely
determined by the random coins and messages selected by all the
players. Then for each $E \in \mathbf{E}$, we have $p_E
=\frac{1}{|\mathbf{M}|} 2^{-r_S-r_R-r_A},$ where $r_S, r_R$ and
$r_A$ denote the length of the random coins of $C_S, C_R$ and $C_A$,
respectively. Similarly, we have $p_{\hat{E}} =
\frac{1}{|\mathbf{M}|} 2^{-r_{\hat{S}}-r_{\hat{R}}-r_{\hat{A}}}.$

As $C_{A2}=\bot$ in this case, it has $r_A = r_{M_A}+ r_{A0} +
r_{A1}$, where $r_{M_A}, r_{A0}, r_{A1}$ denote respectively the
length of $C_{M_A}$, $C_{A0},C_{A1}$. Similarly, it has $r_{\hat{A}}
= r_{M_{\hat{A}}}+ r_{\hat{A}0} + r_{\hat{A}1}$.

Note that $r_{A0} = r_{\hat{A}0} = 1$ and $r_{M_A} = r_{M_{\hat{A}}}
= \lceil \log |\mathbf{M}| \rceil$. By the definition of
$\mathbf{W}_2$, we have that $r_R = r_{\hat{R}}$, $r_S =
r_{\hat{A}1}$ and $r_{A1} = r_{\hat{S}}$. Hence it has $r_S+r_R+r_A
= {r_{\hat{S}}+r_{\hat{R}}+r_{\hat{A}}}$, and then $p_E =
p_{\hat{E}}$ holds. \\[0.2cm]
(ii) Let $\bar{\mathbf{S}}_2 = \mathbf{E} \setminus \mathbf{S}_2$
denote the set of \emph{failed} executions. Since $\hat{E} \in
\bar{\mathbf{S}}_2$ holds for any $E \in \mathbf{S}_2$, and the
one-to-one correspondence of $E$ and $\hat{E}$, we get that
$|\mathbf{S}_2| \leq |\bar{\mathbf{S}}_2|$. The probability that
${\rm {\Pi}}$ fails when $M_A \ne M_S$ can be computed as,
    $$
    \begin{array}{lll}
    &&\hspace{-1.3cm}\Pr[M_R \ne M_S \mid M_S \ne M_A] \\
    &=&\Pr[E \in \bar{\mathbf{S}}_2]\\
    &\geq& \sum_{E \in \mathbf{S}_2}p_{\hat{E}}\\
    &=& \sum_{E \in \mathbf{S}_2}p_{E}\\
    &=& \Pr[M_R = M_S \mid M_S \ne M_A]. \qquad %\hfill\Box
    \end{array}
    $$

\end{proof}

From Claim \ref{cla:2.3} we must have $\Pr[M_R \ne M_S \mid M_A\ne
M_S ] \geq \frac{1}{2}$; hence
$$
\begin{array}{lll}
    &&\hspace{-1.35cm}\Pr[M_R \ne M_S] \\
    &\geq& \Pr[M_R \ne M_S \mid M_S \ne M_A]\Pr[M_S \ne M_A] \\
    &\geq& \frac{1}{2}(1 - \frac{1}{|\mathbf{M}|}). \\
\end{array}
$$

On the other hand, since $\rm \Pi$ is a $\delta$ reliable protocol,
we have $\Pr[M_R \ne M_S] \leq \delta$. It follows that $\delta \geq
\frac{1}{2}(1 - \frac{1}{|\mathbf{M}|})$, which contradicts the
assumption on $\rm \Pi$. %$\hfill\blacksquare$
\end{proof}
\subsection{Impossibility of  $(3, 1)$-Round PD-adaptive $(\varepsilon,
\delta)$-SMT-PD Protocol} \label{sec:31imp}

Theorem \ref{the:ed2} says when the invoker of public channel is
known at the start of the protocol, then $(r, 1)$-round SMT-PD
protocol is impossible. In this section we consider protocols that
allow the invoker of public channel depends on the executions; or
more precisely depends on the random coins of players. We call this
type of SMT-PD protocols \emph{PD-adaptive}.

\begin{definition}
A $(r, r')$-round {\rm SMT-PD} protocol $\rm \Pi$ is called
{\rm{\textsf{PD-adaptive}}} if the invoker of the public channel and
the round of invocation of  the public channel are not specified at
the start but depend on $C_S, C_R, C_A$ and $M_S$.

More specifically, for each round $1 \leq i \leq r$, let player $P
\in \{\cal S, R\}$ be the initiator of the round.  Let
$\mathsf{M}_P^{i-1}$ be the set of all messages received by $P$
during the first $i-1$ rounds and that $\mathsf{M}_S^0 = \{M_S\}$
and $\mathsf{M}_R^0 = \emptyset$. We denote by $P_iX_iY_i
\stackrel{\rm def}{=} f_i(\mathsf{M}_P^{i-1}, C_P)$ the traffic of
round $i$, where $P_i$ denotes the traffic over the public channel,
and $X_i$ and $Y_i$ are the traffic over the two sets of wires, one
all corrupted and one all uncorrupted.

Traffic on the public channel, that is $P_i=\bot$ or $P_i \ne \bot$
is determined by $\mathsf{M}_P^{i-1}$ and $C_P$. Moreover, it must
have $P_j = \bot$ if the public channel has been used $r'$ times
before round $j$.
\end{definition}

\begin{theorem}\label{the:31imp}
Let $n\leq 2t$. Then a {\rm PD-adaptive} $(3, 1)$-round
$(\varepsilon, \delta)$-{\rm SMT-PD} protocol must have
\[3\varepsilon + 2\delta \geq 1 - \frac{3}{|\mathbf{M}|}.\]
\end{theorem}
\begin{proof}
Suppose $\rm \Pi$ is an arbitrarily PD-adaptive $(3, 1)$-round
$(\varepsilon, \delta)$-SMT-PD protocol. We construct a
\emph{static} adversary $\cal A$ that breaks  privacy or reliability
of $\rm \Pi$ and so prove that $3\varepsilon + 2\delta \geq 1 -
\frac{3}{|\mathbf{M}|}$ should hold for any $\rm \Pi$. The message
distribution is assumed to be uniform in this proof.

$\cal A$ selects the first or last $t$ wires to corrupt. In the
rounds before invocation of the public channel, $\cal A$ conducts
man-in-the-middle attack between $\cal S$ and $\cal R$ by tampering
with the corrupted wires. When player $P \in \{\cal S, R\}$   uses
public channel, $\cal A$ simply blocks the corrupted wires and
continues to cheat $P$ by tampering the later transmissions (from
the other player $\bar{P}$ to $P$) over the corrupted wires until
the end of the protocol.

Observe that despite $\bar{P}$ will learn the locations of corrupted
channels, but since the public channel has been used, $\bar{P}$
cannot notify $P$. Thus $\cal A$ can continue to cheat $P$ in the
later execution of the protocol. We will prove that $\cal A$ can
conduct the above attack and thus violate the privacy or reliability
of the protocol.

We use $[A-B-C]$ to indicate the initiators of the first, second and
third rounds are $A$, $B$ and $C$, respectively. The proof is
divided into four steps stated as lemmas, each proving an
impossibility result for an interaction order. The omitted proofs
can be found in Appendix \ref{app:proofs}.

\begin{lemma}\label{lem:sssimp}
If the interaction order of protocol $\rm \Pi$ is $[\cal S-S-S]$,
then $\varepsilon + \delta \geq 1 - \frac{1}{|\mathbf{M}|}$.
\end{lemma}
\begin{proof}
The invoker of public channel in this case must be $\cal S$ and so
$\cal A$ only blocks the traffic over the corrupted wires. This is
an special case of Theorem 2 and we have $\varepsilon + \delta \geq
1 - \frac{1}{|\mathbf{M}|}$. %$\hfill\Box$
\end{proof}

\begin{lemma}\label{lem:srsimp}
If the interaction order of protocol $\rm \Pi$ is $[\cal S-R-S]$,
then $\varepsilon + \delta \geq \frac{1}{2} -
\frac{1}{|\mathbf{M}|}$.
\end{lemma}

\begin{lemma} \label{lem:rrsimp}
If the interaction order of protocol $\rm \Pi$ is $[\cal R-R-S]$,
then $3\varepsilon + 2\delta \geq 1 - \frac{3}{|\mathbf{M}|}$.
\end{lemma}

\begin{lemma} \label{lem:rssimp}
If the interaction order of protocol $\rm \Pi$ is  $[\cal R-S-S]$,
then $\varepsilon + \delta \geq \frac{1}{2} -
\frac{1}{|\mathbf{M}|}$.
\end{lemma}

The above argument shows that a protocol with order $[\cal R-R-S]$
may have better security than protocols with other interaction
orders. However, even in this case, the protocol cannot guarantee
privacy and reliability at the same time. This
completes the proof. %%$\hfill \blacksquare$
\end{proof}

%Notice that the above method does not  \emph{cannot} be used to
%prove the possibility of $(r>3, 1)$-round PD-adaptive protocols.
% We leave it as an open problem.
%
%
\section{An Round Optimal SMT-PD Protocol}\label{sec:0dcons}
As noted earlier the modified version of the protocol in \cite{G08}
has optimal round complexity but has linear (in $n$) transmission
rates over the wires and the public channel, while the complexity of
protocol in \cite{FW00} is similar.

In this section we describe a $(3,2)$-round $(0, \delta)$-SMT-PD
protocol with constant transmission rate over the public channel,
and $O(n)$ transmission rate over the wires (when the message is
long enough).

\subsection{Our Construction}
The proposed protocol uses universal hash functions.
\begin{definition}\label{def:asuhf} Let $m > \ell$. A
function family $\mathcal{H} = \{h: \{0, 1\}^m \to \{0, 1\}^\ell\}$
is called {\textsf{\em{$\gamma$-almost strongly universal$_2$ hash
function family}}} if given any $a_1, a_2 \in \{0, 1\}^m, a_1 \ne
a_2$, and any $b_1, b_2 \in \{0, 1\}^\ell$, it holds that $ \Pr_{h
\in \mathcal{H}}[h(a_1) = b_1 \wedge h(a_2) = b_2] \leq \gamma.$
\end{definition}
\frametext{
\begin{enumerate}
\item ($\mathcal{S}$ $\longrightarrow$ $\mathcal{R}$): For $i = 1,
\ldots, n$, $\mathcal{S}$ randomly selects $r_i \in \{0, 1\}^{\ell}$
and $ R_i \in \{0, 1\}^{m}$ and sends the pair $(r_i, R_i)$ to
$\mathcal{R}$ \emph{along wire $i$}.

\item ($\mathcal{S}$ $\stackrel{P}{\longleftarrow}$ $\mathcal{R}$):
  For $i = 1, \dots, n$, if $\mathcal{R}$ correctly receives a pair $(r_i', R_i')$
  along wire $i$ (i.e., $r_i' \in \{0, 1\}^{\ell}, R_i' \in \{0, 1\}^m$),
  he selects $h_i \from \mathcal{F}$ and
  computes $T_i' = r_i' \oplus h_{i}(R_i')$;
  otherwise, wire $i$ is assumed \emph{corrupted}. He then constructs an
  indicator bit string $B=b_1b_2\cdots b_n$  where $b_i = 1$ if the wire $i$ is
  corrupted and  $b_i = 0$ otherwise. Finally, he sends $(B, (H_1, \dots, H_n))$
  \emph{over the public channel}, where $H_i = (h_i, T_i')$ if
  $b_i=0$; and  $H_i$ is empty, otherwise.

\item ($\mathcal{S}$ $\stackrel{P}{\longrightarrow}$ $\mathcal{R}$):
  $\cal S$ ignores the wires with $b_i = 1$. For $i = 1, \ldots, n$,
  if $b_i = 0$, $\mathcal{S}$ computes $T_i = r_i \oplus h_{i}(R_i)$ and checks $T_i' \overset{?}{=}
  T_i$; if $T_i = T_i'$, wire $i$ is assumed
  \emph{consistent}; otherwise, wire $i$ is corrupted.

  $\mathcal{S}$ constructs an indicator bit string $V=v_1v_2\cdots
v_n$, where $v_i = 1$ if wire $i$ is considered consistent;
otherwise $v_i = 0$. Finally, she publishes the pair $(V, C = M_S
\oplus \{\mathop \oplus \limits_{v_i=1} R_{i}\})$ \emph{over the
public channel}.

\textbf{$\cal{R}$ recovers the message:}  When gets $(V, C)$,
$\mathcal{R}$ recovers $M_R = C \oplus \{\mathop \oplus
\limits_{v_i=1}
  R'_{i}\}$ and outputs it.
\end{enumerate}
} {The $(3,2)$-round $(0, \delta)$-SMT-PD protocol ${\rm \Pi}_1$}
\label{tab:conuhf2}

\begin{corollary} \label{cor:uhf2}
Let $\mathcal{H} = \{h: \{0, 1\}^m \to \{0, 1\}^\ell\}$ be a
$\gamma$-almost strongly universal$_2$ hash function family. Then,
for any $(a_1, c_1) \ne (a_2, c_2) \in \{0, 1\}^m \times \{0,
1\}^\ell$, $\Pr_{h \in \mathcal{H}}[c_1 \oplus h(a_1) = c_2 \oplus
h(a_2)] \leq 2^\ell\gamma.$
\end{corollary}
\begin{proof} For  equality $c_1 \oplus
h(a_1) = c_2 \oplus h(a_2)$, if $a_1 = a_2$, then $c_1 = c_2$ . Thus
we only consider the case of $a_1 \ne a_2$. Since
\begin{eqnarray*}
&&\hspace{-1.0cm}\Pr_{h \in \mathcal{H}}[c_1 \oplus h(a_1) = c_2 \oplus h(a_2)]\\
&=& \sum_{b \in\{0, 1\}^\ell}\Pr_{h \in \mathcal{H}}[h(a_1) = c_1
\oplus b \wedge h(a_2) = c_2 \oplus b].
\end{eqnarray*}
From Definition \ref{def:asuhf}, $\Pr_{h \in \mathcal{H}}[h(a_1) =
c_1 \oplus b \wedge h(a_2) = c_2 \oplus b] \leq \gamma$ and so
$\Pr_{h \in \mathcal{H}}[c_1 \oplus h(a_1) = c_2 \oplus h(a_2)] \leq
2^\ell\gamma,$ and the result follows. %$\hfill\blacksquare$
\end{proof}

Wegman and Carter \cite{WC81} constructed a
$2^{1-2\ell}$-\emph{almost} strongly universal$_2$ hash family
$\mathcal{F} = \{h: \{0, 1\}^{m} \to \{0, 1\}^\ell\}.$ Functions in
$\cal F$ can be described by $O(\ell \log m)$ bits and computed in
polynomial time. The short description length of the family $\cal F$
allows us to authenticate messages with low communication
complexity. The protocol ${\rm \Pi}_1$ transmits $M_S \in \{0,1\}^m$
to $\cal R$ is described in Fig. \ref{tab:conuhf2}. %\vspace{0.2cm}
\begin{theorem} \label{the:uhf}
The protocol ${\rm \Pi}_1$ is a $(3,2)$-round $(0, (n-1)\cdot
2^{1-\ell})$-{\rm SMT-PD} protocol. Moreover, ${\rm \Pi}_1$ is
polynomial time computable, and its transmission rate is $O(n)$ over
the wires and constant over the public channel when
$m=\Omega(n^2\kappa^2)$, where $\kappa$ is the reliability parameter
of the system with $\delta = (n-1)\cdot 2^{1-\ell} = 2^{-\kappa}$.
\end{theorem}
\begin{proof} Let $\mathbf{Cor} = \{i \mid \mbox{wire } i
\mbox{ is corrupted}\}$, and $\mathbf{Con}=\{i \mid \mbox{wire } i
\mbox{ is consistent}\}$.
\begin{itemize}
\item{\textbf{Reliability}}:
If $\cal S$ can detect all corrupted wires with $(r_i', R_i') \ne
(r_i, R_i)$, the protocol is thus perfectly reliable; otherwise, one
such a wire will break the reliability. Using Corollary 2, we show
this probability is small. A more formal proof  follows.

\quad In the second round the wires with $b_i = 1$ are detected as
corrupted, and are ignored in the third round. Hence in the
following we only consider wires with $b_i = 0$. For wire $i$, the
wire is called \emph{bad} if $(r_i, R_i) \ne (r_i', R_i')$ but $r_i
\oplus h_i(R_i) = r_i' \oplus h_i(R_i')$. Bad wires are always
included in $\mathbf{Con}$. Using Corollary \ref{cor:uhf2} and
noting that $r_i, R_i, r_i', R_i'$ are fixed before the second round
and then $h_i$ is  selected with uniform distribution, we have
\begin{eqnarray*}
\Pr[\mbox{wire } i \mbox{ is bad }]&& \\
&&\hspace{-3.3cm} =\  \Pr[r_i \oplus h_i(R_i)= r_i'
\oplus h_i(R_i') \wedge (r_i, R_i) \ne (r_i', R_i')] \\
&&\hspace{-3.3cm}\leq\ \Pr[r_i \oplus h_i(R_i) = r_i' \oplus
h_i(R_i') \mid (r_i, R_i) \ne (r_i', R_i')] \\
&&\hspace{-3.3cm} \leq\ 2^{1-\ell},
\end{eqnarray*}
where the probability is over the random coins of all the players.

\quad Then, the probability of unreliable message transmission is
$$
\begin{array}{lll}
\Pr[M_R \ne M_S] &=& \Pr[\oplus_{j \in \mathbf{Con}}R_j \ne
\oplus_{j
\in \mathbf{Con}}R_j']\\
&\leq& \Pr[\exists j \in \mathbf{Con} \mbox{ s.t. } R_j \ne R_j'] \\
&\leq& \Pr[\exists \mbox{ at least one bad wire}] \\
&\leq& \sum_{j \in \mathbf{Cor}} \Pr[\mbox{wire } j \mbox{ is bad }] \\
&\leq& (n - 1) \cdot 2^{1-\ell},
\end{array}
$$
where the probability is over the random coins of all the players.

\item{\textbf{Perfect Privacy}}: The intuition for proving \emph{perfect privacy}
is as follows:  the adversary can obtain transmissions related to
$M_S$ only from the public channel in round 3. However, $M_S$ is
masked by $R_i$ (if wire $i$ is uncorrupted), and the adversary
knows nothing about $R_i$  because the only transmission which
depends on $R_i$ is in the second round invocation of public channel
($h(R_i)$) which is  masked by $r_i$ and is not known by the
adversary. This is true because $r_i$ was only transmitted on a
secure wire $i$. A more formal proof follows.

\quad Let $M_S = m^*$ be the message chosen by $\cal S$ and $C_A =
c_A$ denotes the value of $\cal A$'s coin.  We first describe
$\mathcal{A}$'s view in the protocol. Observe that in protocol $\rm
\Pi_1$ $\mathbf{Cor}$ is formed completely in the first round since
the last two rounds are only over the public channel. Then in the
first round $\mathcal{A}$ sees $\{(r_i, R_i)\}_{i \in \mathbf{Cor}}$
over the corrupted wires and modifies them into $\{(r_i', R_i')\}_{i
\in \mathbf{Cor}}$. In the second and third round, $\mathcal{A}$
sees respectively $(B, (H_1, \dots, H_n))$ and $(V, M \oplus
\{\oplus R_i\}_{i \in \mathbf{Con}} )$ over the public channel.
Since $\{(r_i', R_i')\}_{i \in \mathbf{Cor}}$ is computed by
$\mathcal{A}$ using $c_A$ and $\{(r_i, R_i)\}_{i \in \mathbf{Cor}}$
(in adaptive way), and when $\mathcal{A}$ knows $\{(r_i', R_i')\}_{i
\in \mathbf{Cor}}$ and $\{h_i\}_{i \in \mathbf{Cor}}$, she can
compute $(\{r_i' \oplus h_i(R_i')\}_{i \in \mathbf{Cor}}, B)$ and
$(\oplus_{i \in \mathbf{Cor}\cap \mathbf{Con} }R_i, V)$ by herself,
we thus remove the computable part from her view and describe it as
a 4-tuple of random variables as follows,
\begin{equation} \label{equ:view}
\begin{array}{lll}
\hspace{-0.2cm}V_A(m^*, c_A) = (c_A, V_1, V_2, V_3)&& \\
=\  (c_A, \{(r_i, R_i)\}_{i \in \mathbf{Cor}}, &&\\
(\{h_i\}_{i=1}^n, \{r_i \oplus h_i(R_i)\}_{i \notin
\mathbf{Cor}}),m^* \oplus (\oplus_{i \notin \mathbf{Cor}} R_i)).&&
\end{array}
\end{equation}
where $V_i$ is $\cal A$'s view in round $i$.

\quad For two messages $m_0, m_1$ and $C_A = c_A$, the statistical
distance between $V_A(m_0, c_A)$ and $V_A(m_1, c_A)$ is given by,
$$
\begin{array}{lll}
\hspace{-0.1cm}{\rm \Delta}(V_A(m_0, c_A), V_A(m_1, c_A))&&\\
= \frac{1}{2}\sum_v \mid \Pr[V_A(m_0, c_A)=v] -\Pr[V_A(m_1,
c_A)=v]\mid,&&
\end{array}
$$
where the probability is over the choices of $C_S$ and $C_R$. Then
the term $\Pr[V_A(m_0, c_A)=v]$ is given by,
$$
\begin{array}{lll}
\hspace{-0.1cm}\Pr[V_A(m_0, c_A)=v]&& \\
=\ \sum_{\{c_S, c_R:  V_A(m_0, c_A)=v\}} \Pr[C_S=c_S \wedge
C_R=c_R].&&
\end{array}
$$

\quad Note that $C_S$ and $C_R$ are independent and have length
$n(m+\ell)$ and $wk$ respectively, where $w$ is the Hamming weight
of the string $B$ and $k$ is the description length of function in
$\cal F$. Hence $\Pr[C_S=c_S \wedge C_R=c_R] =
\frac{1}{2^{n(m+\ell)+wk}}$; note this value is independent of the
value of $m_0$.

\quad Therefore we only need to count the number of executions in which the coin
tosses of the sender and the receiver are such that random variable
$V_A(m_0, c_A)=v$.

\quad Suppose that $v= (c_A,V_1,V_2,V_3)$ is fixed, it implies that $\mathbf{Cor}$
and $c_R = \{h_i\}_{i=1}^n$ are also determined; then the choices of
$\{(r_i, R_i)\}_{i \notin \mathbf{Cor}}$ should be consistent with
$V_2$ and $V_3$. Since $\oplus_{i \notin \mathbf{Cor}}R_i = V_3
\oplus m_0$, when $m_0, V_3$ are fixed, at most $n - |\mathbf{Cor}|
- 1$ elements in $\{R_i\}_{i \notin \mathbf{Cor}}$ can be selected
freely. Moreover, when $V_2$ and $\{R_i\}_{i \notin \mathbf{Cor}}$
are fixed, $\{r_i\}_{i \notin \mathbf{Cor}}$ are also determined.
Therefore, the number of $C_S, C_R$ result in $V_A(m_0, c_A)=v$ are
bounded by the number of $R_i$ for $i \notin \mathbf{Cor}$. Totally,
they have $2^{m  (n- \mid \mathbf{Cor}\mid -1)}$ different choices.
Hence we have,
\[
\Pr[V_A(m_0, c_A)=v] = \frac{2^{m(n- \mid \mathbf{Cor}\mid
-1)}}{2^{n(m+\ell)+wk}}.
\]
\quad The proof is complete by noting that the above probability is
independent of  $m_0$. %$\hfill\Box$
\item{\textbf{Complexity}}: Since the hash function is
polynomial time computable in $m$, the \emph{computation complexity}
of $\mathcal{S}$ and $\mathcal{R}$ are  polynomial in $n$ and $m$.
For \emph{communication complexity}, ${\rm \Pi}_1$ needs to
communicate $m + \ell$ bits over each wire, and at most $(4s \log m
+ \ell+2)n + m$ bits over the public channel, where $s = \ell +
\log\log m$. If the reliability requirement is set to $\delta =
2^{-\kappa}=(n - 1) \cdot 2^{1-\ell}$, then $\ell = \kappa + \log
(n-1) + 1$. The \emph{transmission rate} over the public channel
assuming $m=\Omega(n^2\kappa^2)$, is
 $((4s\log m + \ell + 2)n+m) / m$ which is constant asymptotically.%$\hfill\blacksquare$
\end{itemize}
\end{proof}

\subsection{Comparisons with Schemes in
\cite{FW00,G08}}\label{sec:com}
%\begin{enumerate}

As noted earlier communication over public channel is much more
costly than communication over wires, and so minimizing the
transmission rate over the public channel will have a large effect
on  overall efficiency of the protocol. This is particularly
important for transmitting long messages. For example in most cases
$\kappa=30$ provide sufficient reliability. However messages can be
as long as $2^{20}$ bits. When $n=30$ wires are available, our
proposed protocol transmits around $2^{20}$ bits over the public
channel with reliability higher than $1- 2^{-30}$ (since $m >
n^2\kappa^2$). The protocols in \cite{FW00,G08} both have
transmission rate $O(n)$ and so need to send almost $30 $ times data
($30 \times 2^{20} \approx 2^{25}$ bits) over the public channel.
The reliability is $1-2^{-O(m)} = 1- 2^{-2^{20}}$ in
\cite{FW00,G08}, which would be unnecessarily high.
%\end{enumerate}
\section{Conclusion and Further Research}
In this work we considered  round optimality protocols for secure
message transmission (SMT) by public discussion. This is an
important communication model in realizing almost-everywhere
multiparty computation. Since the implementation cost of public
channel is high, it is important to minimize transmission over the
pubic channel.   Our results show that secure protocol in this model
need at least 3 rounds  and in 2 of them the public channel must be
invoked. We prove this result in a general setting where the
invocation of public channel is not known at the start of the
protocol and depends on the coin tosses of participants.   We
describe a round optimal protocol that has \emph{constant}
transmission rate over the public channel and linear transmission
rate over other wires.

Existence of PD-adaptive SMT-PD protocols with $r \geq 4$ rounds and
one round public discussion, and construction of round optimal
protocols with optimal communication complexity over wires and
public channel (if there exists) are interesting open problems.

\begin{appendix}
\subsection{Proof for Lemma \ref{cor:advdis}}
\label{cor:advdisprf}
\begin{proof}
\quad By Definition \ref{def:smt} and Lemma \ref{lem:sdci} we have:
For any algorithm $\mathcal{D}$, any two messages $m_0, m_1 \in
\mathbf{M}$, and any adversary $\cal B$ with randomness $c_B \in
\{0, 1\}^*$,
\begin{equation}\label{eq:advdisfull}
|\Pr[{\cal D}(V_B(m_0,c_B))= 1] - \Pr[{\cal D}(V_B(m_1,c_B))= 1]|
\leq \varepsilon,
\end{equation} where the
probability is over the random coins of $\mathcal{S}$ and
$\mathcal{R}$. Note here $V_B(m, c)$ is (the random variable of) the
view of $\cal B$ when the (fixed) message $m \in \mathbf{M}$ is
transmitted and $\cal B$ uses the (fixed) coins $C_B = c_B$ in the
protocol.

Then by taking \emph{average} over the randomness of $C_B$, the
following holds from Eq.(\ref{eq:advdisfull})
\begin{equation} \label{eq:advexp}
|\Pr[{\cal D}(V_B(m_0))= 1] - \Pr[{\cal D}(V_B(m_1))= 1]| \leq
\varepsilon \,,
\end{equation}
where $V_B(m)$ denotes the view of $\cal B$ when the fixed message
$m \in \mathbf{M}$ is transmitted in the protocol, and it is a
random variable over the random coins of $\cal S, R$ and $\cal B$.

The adversary's strategy consists of: selecting messages $(M_0, M_1)$ followed by
 attacking the protocol and so we write
$\mathcal{B}=(\mathcal{B}_1, \mathcal{B}_2)$. We use $C_{B1}$ to
denote the random coins used by $\mathcal{B}_1$ to select $(M_0,
M_1)$. Let $p_0 \stackrel{\rm def}{=} \Pr[\mathcal{B}_2^{{\rm
\Pi}(m_0)}()=1]$ and $p_1 \stackrel{\rm def}{=}
\Pr[\mathcal{B}_2^{{\rm \Pi}(m_1)}()=1]$. We have,
$$
\begin{array}{lll}
&&\hspace{-1.0cm}\left|\Pr\left[\mathcal{B}^{{\rm \Pi}(M_0)}()=1
\right] -
\Pr\left[\mathcal{B}^{{\rm \Pi}(M_1)}()=1 \right]\right| \\[0.2cm]
&=&\left|\sum_{C_{B1}=c} \Pr[C_{B1} =
c]\left( p_0 - p_1 \right)\right|\\[0.2cm]
&\leq& \sum_{C_{B1}=c} \Pr[C_{B1} =
c]\left|p_0 - p_1 \right| \\
&\leq& \varepsilon \,.
\end{array}
$$
The last step follows from the observation that
$|p_0-p_1| \leq \varepsilon$ due to (\ref{eq:advexp}). %$\hfill\Box$
\end{proof}

\subsection{Proofs Omitted From Theorem
\ref{the:31imp}}\label{app:proofs}

As in the proof of Theorem \ref{the:ed2}, we separate $\cal A$'s
random coins into four parts: $(C_{M_A}, C_{A0}, C_{A1}, C_{A2})$.
For the sake of clarity, the message \emph{selected} by $\cal A$
using randomness $C_{M_A}$ is denoted by $M_A$, while the message
\emph{outputted} by $\cal A$ by the end of the protocol is denoted
by $M_A^+$.
\subsubsection{Proof of Lemma \ref{lem:srsimp}} \mbox{  }\\
%\begin{proof}
The public channel can be used in any of the three rounds. For
simplicity, we assume $C_{A0} = 1$, i.e., $\cal A$ selects the last
$t$ wires to corrupt. The actions of $\cal A$ is illustrated as in
Fig. \ref{fig:srsE1}, \ref{fig:srsE2} and \ref{fig:srsE3}
respectively. (We remark that when $C_{A0} = 0$, $\cal A$'s action
is similar.) The detail of $\cal A$ selecting $(M_A, C_{A1},
C_{A2})$ when $\cal S$ doesn't use the public channel in the first
round is supplied in Fig. \ref{fig:Amacasrs}.

We remark that: (i) When $\cal S$ doesn't use public channel in
round 1 and $\Omega_2 \ne \emptyset$, the strategy as described in
Fig. \ref{fig:Amacasrs} ensures that $\cal A$ can produce message
$X_2'Y_2'$ without public channel communication in the second round.
(ii) Since $\cal A$ is computationally unbounded, she knows $f_1$
and $f_2$'s function tables and so knows the sets $\Omega_1$ and
$\Omega_2$. Thus $\cal A$ can conduct the above attacks.

We analyze the success probability of $\cal A$ in the following. Let
$\mathcal{E}_1$ and $\mathcal{E}_3$ denote the events that $\cal S$
invokes the public channel in round 1 and 3, respectively. Let
$\mathcal{E}_2$ be the event that $\cal R$ invokes the public
channel in round 2. Then $\mathcal{E}_1$, $\mathcal{E}_2$ and
$\mathcal{E}_3$ are disjoint events and $\Pr[\mathcal{E}_1 \vee
\mathcal{E}_2 \vee \mathcal{E}_3] = 1$ since $\rm \Pi$ is a
$(3,1)$-round protocol.

\frametext{ Assume in the first round $\cal S$ sends   $X_1Y_1$ and
let the sets $\Omega_1 \subseteq \mathbf{M} \times \{0,1\}^*$ and
$\Omega_2 \subseteq \Omega_1 \times \{0,1\}^*$ be defined as
$$
\begin{array}{lll}
\Omega_1 &\stackrel{\rm def}{=}& \{(m, c_1) \mid f_1(m, c_1) \mbox{
doesn't use}\\
&& \mbox{ public channel }\}\end{array}
$$
and
$$
\begin{array}{lll}
\Omega_2 &\stackrel{\rm def}{=}& \{(m, c_1, c_2) \mid (m, c_1)\in
\Omega_1, c_2 \in \{0, 1\}^* \\
&& \mbox{s.t. } f_2(X_1'Y_1, c_2) \mbox{ doesn't use public}\\
&& \mbox{ channel where } X_1'Y_1' = f_1(m, c_1)\}.
\end{array}
$$
We have $(M_S, C_S) \in \Omega_1$. If $\Omega_2 \ne \emptyset$,
$\cal A$ randomly chooses $(M_A, C_{A1}, C_{A2}) \from \Omega_2$;
otherwise, $\cal A$ randomly chooses $(M_A, C_{A1}, C_{A2}) \from
\Omega_1 \times \{0, 1\}^*$. }{The strategy that $\cal A$ selects
$(M_A, C_{A1}, C_{A2})$ when $\cal S$ doesn't use public channel in
round 1.} \label{fig:Amacasrs}

\begin{figure*}[!ht]
$$
\input xy
\xyoption{all}  \xymatrix@R=0.05in@C=1.1in {
\mathcal{\underline{S}}_{\quad(M_S, C_S)}&
\mathcal{\underline{A}}_{\quad \left( C_{A0},C_{A2} \right)}&
\mathcal{\underline{R}}_{\quad(C_R)} \\
\txt<7pc>{$P_1X_1Y_1=f_1(M_S, C_S)$}
\ar[r]^-{P_1X_1Y_1}&*\txt<8pc>{blocks $Y_1$}
\ar[r]^-{P_1X_1}&*\txt<8pc>{} \\
\txt<7pc>{} &\ar[l]_-{\large X_2Y_2'}
\txt<8pc>{$X_2'Y_2'=f_2(P_1Y_1,C_{A2})$}
&\ar[l]_-{\large X_2Y_2} \txt<7pc>{$X_2Y_2=f_2(P_1X_1, C_R)$} \\
\txt<7pc>{$X_3Y_3=f_3(M_S, X_2Y_2', C_S)$}
\ar[r]^-{X_3Y_3}&*\txt<8pc>{blocks $Y_3$, \\computes $M_A^+ =
g(P_1Y_1,Y_3,C_{A2})$}
\ar[r]^-{X_3}&*\txt<8pc>{$M_R = g(P_1X_1, X_3, C_R)$} \\
}
$$
\caption{An execution of $\rm \Pi$ with order $[\cal S-R-S]$, where
$C_{A0}=1$ and $\cal S$ uses the public channel in round
1.}\label{fig:srsE1}

$$
\input xy
\xyoption{all}  \xymatrix@R=0.05in@C=1.1in {
\mathcal{\underline{S}}_{\quad(M_S, C_S)}&
\mathcal{\underline{A}}_{\quad \left( C_{A0} \right)}&
\mathcal{\underline{R}}_{\quad(C_R)} \\
\txt<7pc>{$X_1Y_1=f_1(M_S, C_S)$}
\ar[r]^-{X_1Y_1}&*\txt<8pc>{selects $(M_A, C_{A1}, C_{A2})$,
\\ $X_1'Y_1' = f_2(M_A, C_{A1})$}
\ar[r]^-{X_1Y_1'}&*\txt<7pc>{} \\
\txt<7pc>{} &\ar[l]_-{\large P_2X_2} \txt<8pc>{blocks $Y_2$}
&\ar[l]_-{\large P_2X_2Y_2} \txt<7pc>{$P_2X_2Y_2=f_2(X_1Y_1', C_R)$} \\
\txt<7pc>{$X_3Y_3=f_3(M_S, P_2X_2, C_S)$}
\ar[r]^-{X_3Y_3}&*\txt<8pc>{$X_3'Y_3' = f_3(M_A, P_2Y_2,C_{A1})$}
\ar[r]^-{X_3Y_3'}&*\txt<8pc>{$M_R = g(X_1X_1', X_3Y_3', C_R)$} \\
}
$$
\caption{An execution of $\rm \Pi$ with order $[\cal S-R-S]$, where
$C_{A0}=1$ and $\cal R$ uses the public channel in round
2.}\label{fig:srsE2}

$$
\input xy
\xyoption{all}  \xymatrix@R=0.05in@C=1.1in {
\mathcal{\underline{S}}_{\quad(M_S, C_S)}&
\mathcal{\underline{A}}_{\quad \left( C_{A0} \right)}&
\mathcal{\underline{R}}_{\quad(C_R)} \\
\txt<7pc>{$X_1Y_1=f_1(M_S, C_S)$}
\ar[r]^-{X_1Y_1}&*\txt<8pc>{selects $(M_A, C_{A1}, C_{A2})$,
\\ $X_1'Y_1' = f_2(M_A, C_{A1})$}
\ar[r]^-{X_1Y_1'}&*\txt<8pc>{} \\
\txt<7pc>{} &\ar[l]_-{\large X_2Y_2'}
\txt<7pc>{$X_2'Y_2'=f_2(X_1'Y_1,C_{A2})$}
&\ar[l]_-{\large X_2Y_2} \txt<7pc>{$X_2Y_2=f_2(X_1Y_1', C_R)$} \\
\txt<7pc>{$P_3X_3Y_3=f_3(M_S, X_2Y_2', C_S)$}
\ar[r]^-{P_3X_3Y_3}&*\txt<8pc>{blocks $Y_3$, \\computes  $M_A^+ =
g(X_1'Y_1, P_3Y_3, C_{A2})$}
\ar[r]^-{P_3X_3}&*\txt<8pc>{$M_R = g(X_1Y_1', P_3X_3, C_R)$} \\
}
$$
\caption{An execution of $\rm \Pi$ with order $[\cal S-R-S]$, where
$C_{A0}=1$ and $\cal S$ uses the public channel in round
3.}\label{fig:srsE3}
\end{figure*}

\begin{myclaim}\label{cla:mamseq}
Let $b \in \{1, 3\}$. If $\mathcal{E}_b$ occurs, we have
\[\Pr[M_A^+=M_S \mid \mathcal{E}_b] \geq \Pr[M_R=M_S \mid
\mathcal{E}_b].\]
\end{myclaim}
\begin{proof}
(i) We first prove the case of $b=1$. Denote by $\mathbf{E}_1$ the
set of all executions where $\mathcal{E}_1$ occurs, and by
$\mathbf{S}_1 \subseteq \mathbf{E}_1$ the set of successful
executions in which $\cal R$ outputs $M_R = M_S$.

Define a relation $\mathbf{W}_1 \subseteq \mathbf{E}_1 \times
\mathbf{E}_1$, where $(E, \hat{E}) \in \mathbf{W}_1$ if: (i) $M_S,
C_S$ remain unchanged in the two executions; (ii) $C_{\hat{A}0}
\oplus C_{A0} = 1$; (iii) $C_{A2} = C_{\hat{R}}, C_R =
C_{\hat{A}2}$.

Similar to CASE 1 in Theorem 2, we can prove that $\cal S$ cannot
distinguish two swapped executions $(E, \hat{E}) \in \mathbf{W}_1$
and so if $M_R = M_S$, we have $M_{\hat{A}}^+ = M_S$. Furthermore,
we have $p_E = \frac{1}{|{\rm \Phi}|}2^{-r_A-r_R} = p_{\hat{E}}$,
where ${\rm \Phi} \subseteq \mathbf{M} \times \{0,1\}^{r_S}$ is the
set of all $(M_S, C_S)$ such that $\mathcal{E}_1$ occurs, and $r_S,
r_A, r_R$ denote the length of the randomness used by $\cal S, A,
R$, respectively. We then obtain,

$$
\begin{array}{lll}
\Pr[M_A^+=M_S \mid \mathcal{E}_1] &\geq& \sum_{E \in \mathbf{S}_1}p_{\hat{E}}\\
&=& \sum_{E \in \mathbf{S}_1}p_E \\
&=& \Pr[M_R=M_S \mid \mathcal{E}_1].
\end{array}
$$

\noindent (ii) When $b=3$, let $\mathbf{E}_3$ be the set of all
executions where $\mathcal{E}_3$ occurs, and $\mathbf{S}_3 \subseteq
\mathbf{E}_3$ be the set of all successful executions in which $\cal
R$ outputs $M_R = M_S$. Define a relation $\mathbf{W}_3 \subseteq
\mathbf{E}_3 \times \mathbf{E}_3$, where $(E, \hat{E}) \in
\mathbf{W}_3$ if: (i) $M_S, C_S$ and $M_A, C_{A1}$ remain unchanged
in the two executions; (ii) $C_{\hat{A}0} \oplus C_{A0} = 1$; (iii)
$C_{A2} = C_{\hat{R}}, C_R = C_{\hat{A}2}$.

Then by a similar proof of CASE 1 in Theorem 2, we have
$M_{\hat{A}}^+ = M_R$.

For any two executions $(E, \hat{E}) \in \mathbf{W}_3$, suppose
$(M_S, C_S, C_R, C_A)= (m_S, c_S, c_R, c_A)$ and $(M_{\hat{S}},
C_{\hat{S}}, C_{\hat{R}}, C_{\hat{A}})= (m_{\hat{S}}, c_{\hat{S}},
c_{\hat{R}}, c_{\hat{A}})$. Then the probability that $E$ occurs is
$p_E = \Pr[(M_S, C_S) = (m_S, c_S) \wedge C_R = c_R \wedge C_A = c_A
\mid \mathcal{E}_3] = \alpha\cdot \beta,$ where $\alpha = \Pr[(M_S,
C_S) = (m_S, c_S)\mid \mathcal{E}_3]$ and $\beta= \Pr[C_A = c_A
\wedge C_R = c_R \mid (M_S, C_S) = (m_S, c_S) \wedge
\mathcal{E}_3]$. Similarly, it has $p_{\hat{E}} = \Pr[(M_{\hat{S}},
C_{\hat{S}}) = (m_{\hat{S}}, c_{\hat{S}}) \wedge C_{\hat{R}} =
c_{\hat{R}} \wedge C_{\hat{A}} = c_{\hat{A}} \mid \mathcal{E}_3] =
\hat{\alpha}\cdot \hat{\beta},$ where $\hat{\alpha}=
\Pr[(M_{\hat{S}}, C_{\hat{S}}) = (m_{\hat{S}}, c_{\hat{S}})\mid
\mathcal{E}_3]$ and $\hat{\beta}= \Pr[C_{\hat{A}} = c_{\hat{A}}
\wedge C_{\hat{R}} = c_{\hat{R}} \mid (M_{\hat{S}}, C_{\hat{S}}) =
(m_{\hat{S}}, c_{\hat{S}}) \wedge \mathcal{E}_3]$.

Obviously, it has  $\alpha = \hat{\alpha}$ as $(M_S, C_S) =
(M_{\hat{S}}, C_{\hat{S}})$. The following is to prove $\beta =
\hat{\beta}$. Since $(M_A, C_{A1}) = (M_{\hat{A}}, C_{\hat{A}1})$,
this is equivalent to proving
\begin{equation}\label{equ:equlrdm}
\begin{array}{lll}
&&\hspace{-1.0cm}\Pr[C_{A2} = c_{A2} \wedge C_R = c_R \mid
\mathcal{X}] \\
&=&\ \Pr[C_{\hat{A}2} = c_{\hat{A}2} \wedge C_{\hat{R}} =
c_{\hat{R}} \mid \hat{\mathcal{X}}],
\end{array}
\end{equation}
where $\mathcal{X}$ denotes the event that $(M_S, C_S) = (m_S, c_S)
\wedge (M_A, C_{A0}, C_{A1}) = (m_A, c_{A0}, c_{A1}) \wedge
\mathcal{E}_3$, and $\hat{\mathcal{X}}$ denotes the event that
$(M_{\hat{S}}, C_{\hat{S}}) = (m_{\hat{S}}, c_{\hat{S}}) \wedge
(M_{\hat{A}}, C_{\hat{A}0}, C_{{\hat{A}}1}) = (m_{\hat{A}},
c_{\hat{A}0}, c_{{\hat{A}}1}) \wedge \mathcal{E}_3$.

Note that $C_R$ is uniformly selected by $\cal R$ and $C_{A2}$ is
selected by $\cal A$ in the first round without seeing any
information about $C_R$. Hence $C_{A2}$ and $C_R$ are  independent.
Similarly, $C_{\hat{A}2}$ and $C_{\hat{R}}$ are independent.

Then Eq.(\ref{equ:equlrdm}) can be expressed as
\begin{eqnarray*}
&&\hspace{-1.1cm}\Pr[C_{A2} = c_{A2} \mid \mathcal{X}]\Pr[C_R = c_R
\mid
\mathcal{X}] \\
&=&\Pr[C_{\hat{A}2} = c_{\hat{A}2} \mid
\hat{\mathcal{X}}]\Pr[C_{\hat{R}} = c_{\hat{R}} \mid
\hat{\mathcal{X}}].
\end{eqnarray*}

Let ${\rm \Phi} = \{c \mid f_2(X_1'Y_1, c)\mbox{ doesn't use
public}$ $\mbox{ channel}\}$; where $X_1'Y_1$ comes from
$X_1Y_1=f_1(m_S, c_S)$ and $X_1'Y_1'=f_1(m_A, c_{A1})$. Since
$C_{A2}$ is uniformly selected from ${\rm \Phi}$, we have $
\Pr[C_{A2} = c_{A2} \mid \mathcal{X}] = \frac{1}{|{\rm \Phi}|}. $
Furthermore, when $\hat{\mathcal{X}}$ occurs, from the definition of
$\mathbf{W}_3$ we have that $C_{\hat{R}}$ is in ${\rm \Phi}$, which
implies $\Pr[C_{\hat{R}} = c_{\hat{R}} \mid \hat{\mathcal{X}}] =
\frac{1}{|{\rm \Phi}|}$. Similarly, we get
\[
\Pr[C_R = c_R \mid \mathcal{X}]=\Pr[C_{\hat{A}2} = c_{\hat{A}2} \mid
\hat{\mathcal{X}}].
\]

We thus prove the equality of Eq.(\ref{equ:equlrdm}), which implies
that $p_E = p_{\hat{E}}$, and then
$$
\begin{array}{lll}
\Pr[M_A^+=M_S \mid \mathcal{E}_3] &\geq& \sum_{E \in \mathbf{S}_3}p_{\hat{E}}\\
&=& \sum_{E \in \mathbf{S}_3}p_E \\
&=& \Pr[M_R=M_S \mid \mathcal{E}_3].\qquad %\hfill\Box
\end{array}
$$
\end{proof}

\begin{myclaim}\label{cla:mrmseq}
$\Pr[M_R \ne M_S \mid M_S \ne M_A \wedge \mathcal{E}_2] \geq \Pr[M_R
= M_S \mid M_S \ne M_A \wedge \mathcal{E}_2]$.
\end{myclaim}
\begin{proof}
Denote by $\mathbf{E}_2$ the set of all executions where
$\mathcal{E}_2$ occurs. Let $\mathbf{S}_2 \subseteq \mathbf{E}_2$
denote the set of executions in which $\cal{R}$ outputs $M_R = M_S$
given that $M_A \ne M_S$.

We define a relation $\mathbf{W}_2 \subseteq \mathbf{E}_2 \times
\mathbf{E}_2$ such that $(E, \hat{E}) \in \mathbf{W}_2$ if: (i)
$C_R$ remains unchanged in the two executions; (ii) $C_{\hat{A}0}
\oplus C_{A0} = 1$; (iii) $C_{A1} = C_{\hat{S}}, C_S =
C_{\hat{A}1}$; and (iv) $M_S = M_{\hat{A}}, M_A = M_{\hat{S}}$.

Then $\cal R$ cannot distinguish  two swapped executions $(E,
\hat{E})$ in $\mathbf{W}_2$ and if $E \in \mathbf{S}_2$, we have
$\hat{E} \notin \mathbf{S}_2$. Moreover, for any $E \in
\mathbf{E}_2$, a  proof similar to case (ii) in Claim
\ref{cla:mamseq} can be used to prove that $p_E =p_{\hat{E}}$. We
thus have,
    $$
    \begin{array}{lll}
    &&\hspace{-1.1cm}\Pr[M_R \ne M_S \mid M_S \ne M_A \wedge \mathcal{E}_2] \\
    &=& \Pr[E \notin
    \mathbf{S}_2]\\
    &\geq& \sum_{E \in \mathbf{S}_2}p_{\hat{E}}\\
    &=& \sum_{E \in \mathbf{S}_2}p_{E}\\
    &=& \Pr[M_R = M_S \mid M_S \ne M_A \wedge \mathcal{E}_2]. \qquad %\hfill\Box
    \end{array}
    $$
\end{proof}

From Claim \ref{cla:mamseq} and \ref{cla:mrmseq}, we have
\begin{equation}\label{eq:maems}
\begin{array}{lll}
&&\hspace{-1.1cm}\Pr[M_A^+ = M_S]\\
&\geq& \Pr[M_A^+ = M_S \mid \mathcal{E}_1]\Pr[\mathcal{E}_1] \\
&& + \Pr[M_A^+ = M_S \mid \mathcal{E}_3]\Pr[\mathcal{E}_3] \\
&\geq& \Pr[M_R = M_S \wedge \mathcal{E}_1] + \Pr[M_R = M_S \wedge
\mathcal{E}_3] \,
\end{array}
\end{equation}
and
\begin{equation}\label{eq:msnemr}
\begin{array}{lll}
&&\hspace{-1.1cm}\Pr[M_R \ne M_S] \\
&\geq& \Pr[M_R \ne M_S \mid \mathcal{E}_2]\Pr[\mathcal{E}_2] \\
&\geq& \Pr[M_R \ne M_S \mid M_S \ne M_A \wedge \mathcal{E}_2]\\
&& \cdot\Pr[M_S\ne M_A \mid \mathcal{E}_2]\Pr[\mathcal{E}_2]\\
&\geq& \Pr[M_R = M_S \mid M_S \ne M_A \wedge \mathcal{E}_2]\\
&& \cdot\Pr[M_S\ne M_A \wedge \mathcal{E}_2] \\
&=& \Pr[M_R=M_S \wedge \mathcal{E}_2]\\
&& \cdot(1 - \Pr[M_S = M_A \mid M_R = M_S \wedge \mathcal{E}_2]) \\
&\geq& \Pr[M_R=M_S \wedge \mathcal{E}_2] - \Pr[M_A = M_S]
\end{array}
\end{equation}

Moreover, we also have $\Pr[M_A = M_S] \leq \varepsilon +
\frac{1}{|\mathbf{M}|}$, as otherwise by choosing $M_A^+$ to be
$M_A$, we have $\Pr[M_A^+ = M_S]
> \varepsilon + \frac{1}{|\mathbf{M}|}$, which contradicts  Lemma
\ref{lem:edide}.

Hence, it has
$$
\begin{array}{lll}
&&\hspace{-1.1cm} \Pr[M_A^+ = M_S] + \Pr[M_R \ne M_S]\\
&\geq& \Pr[M_R = M_S \wedge
\mathcal{E}_1] + \Pr[M_R = M_S \wedge \mathcal{E}_3]\\
&& + \Pr[M_R = M_S \wedge \mathcal{E}_2] - \Pr[M_A = M_S]\\
&=& \Pr[M_R = M_S] - \Pr[M_A = M_S].
\end{array}
$$
Thus, by noting that $\Pr[M_A^+ = M_S] \leq \varepsilon +
\frac{1}{|\mathbf{M}|}$, $\Pr[M_A = M_S]\leq \varepsilon +
\frac{1}{|\mathbf{M}|}$ and $\Pr[M_S \ne M_R] \leq \delta$, we get
$\varepsilon + \delta \geq \frac{1}{2}-\frac{1}{|\mathbf{M}|}$ \,.
$\hfill \blacksquare$
%\end{proof}
\\
\subsubsection{Proof of Lemma \ref{lem:rrsimp}}
%\begin{proof}
Assume $C_{A0}=1$, we illustrate $\cal A$'s strategy as follows.

\emph{Round 1}: (i) if $\cal R$ uses public channel, $\cal A$ just
blocks the $t$ corrupted wires. Then $\cal A$ selects $(M_A, C_{A1})
\from \mathbf{M} \times \{0, 1\}^*$, and sets $C_{A2}=\bot$.

(ii) Otherwise, assume $\cal R$ sends out $X_1Y_1$. Consider the
following two sets
$$
\begin{array}{lll}
\Omega_1 &\stackrel{\rm def}{=}& \{c \mid c \in \{0,1\}^* \mbox{
s.t. } f_1(c) \mbox{ involves no public} \\
&& \mbox{ channel communication}\}, \\
\Omega_2 &\stackrel{\rm def}{=}& \{c \mid c \in \Omega_1 \mbox{ s.t.
} f_2(c) \mbox{ involves no public}\\
&& \mbox{ channel communication}\}.
\end{array}
$$
Obviously, $C_R \in \Omega_1$. Then if $|\Omega_2| > 0$, $\cal A$
selects $C_{A2} \from \Omega_2$; otherwise, selects $C_{A2} \from
\Omega_1$. $\cal A$ also chooses $(M_A, C_{A1}) \from \mathbf{M}
\times \{0, 1\}^*$, then computes $X_1'Y_1' = f_1(C_{A2})$ and
replaces $Y_1$ by $Y_1'$.

\emph{Round 2}: (i) if $\cal R$ uses public channel in this round or
public channel has been used in round 1, $\cal A$ just blocks the
corrupted wires. (ii) Otherwise, suppose $\cal R$ responses
$X_2Y_2$, it has $C_R \in \Omega_2$, then the selection of $C_{A2}$
ensures that $\cal A$ can produce message $X_2'Y_2'$ without public
channel communication. $\cal A$ thus replaces $Y_2$ by $Y_2'$.

\emph{Round 3:} (i) If $\cal S$ sends out $P_3X_3Y_3$, $\cal A$ just
blocks $Y_3$, and computes $M_A^+ = g(P_3Y_3, C_{A2})$. (ii)
Otherwise, assume $\cal S$ sends out $X_3Y_3$, it implies that
public channel has been used in the first two rounds, $\cal A$ thus
computes $X_3'Y_3'$ and replaces $Y_3$ by $Y_3'$.

Then by a similar calculation of Eq. (\ref{eq:maems}) and
(\ref{eq:msnemr}), we get
\begin{eqnarray*}
&&\hspace{-1.1cm}\Pr[M_R \ne M_S] \\
&\geq& \Pr[M_R=M_S \wedge \mathcal{E}_1]+ \Pr[M_R=M_S \wedge \mathcal{E}_2] \\
&& - 2\Pr[M_S = M_A]
\end{eqnarray*}
and
\begin{eqnarray*}
\Pr[M_A^+ = M_S] &\geq&
\Pr[M_A^+ = M_S \wedge \mathcal{E}_3] \\
&\geq& \Pr[M_R = M_S \wedge \mathcal{E}_3],
\end{eqnarray*}
where $\mathcal{E}_1, \mathcal{E}_2$ denote the events that $\cal R$
uses the public channel in round 1 and 2 respectively, and
$\mathcal{E}_3$ denotes the event that $\cal S$ uses the public
channel in round 3. Finally we obtain $3\varepsilon + 2\delta \geq 1
- \frac{3}{|\mathbf{M}|}$. $\hfill \blacksquare$
%\end{proof}
\mbox{ }\\
\subsubsection{Proof of Lemma \ref{lem:rssimp}}\mbox{ }\\
%\begin{proof}
$\cal A$'s strategy with $C_{A0}=1$ is described as follows.

\emph{Round 1}: (i) If $\cal R$ uses public channel, $\cal A$ just
blocks the $t$ corrupted wires; (ii) otherwise, assume $\cal R$
sends out $X_1Y_1$, $\cal A$ selects $C_{A2}$ from the set of
\begin{eqnarray*}
\Omega_1 &\stackrel{\rm def}{=}& \{c \mid c \in \{0,1\}^* \mbox{
s.t. } f_1(c) \mbox{ involves no public} \\
&& \mbox{ channel communication}\}
\end{eqnarray*} and computes $X_1'Y_1' =
f_1(C_{A2})$, then replaces $Y_1$ by $Y_1'$.

\emph{In the latter two rounds}: (i) If $\cal R$ does \emph{not} use
the public channel in round 1, it says $\cal S$ will be the invoker
of public channel, thus $\cal A$ just blocks the corrupted wires.
(ii) Otherwise, $\cal A$ chooses $(M_A, C_{A1}) \from \mathbf{M}
\times \{0, 1\}^*$ and computes $X_2'Y_2'$ and $X_3'Y_3'$, then
modifies the corrupted wires.

We note that the impossibility proof in this scenario is similar to
Lemma \ref{lem:srsimp}, and thus omit it here. $\hfill \blacksquare$
%\end{proof}

\end{appendix}
\end{document}